# First-Passage Time for Upper Bounds on Fluctuations of Trajectory Observables

V. V. Ryazanov

Institute for Nuclear Research, pr. Nauki, 47 Kiev, Ukraine, e-mail: vryazan19@gmail.com

General upper bounds on fluctuations of trajectory observables were recently obtained. It turned out that the size of fluctuations of dynamical observable is limited from below and from above. For the moment generating function of general upper bounds on the size of fluctuations, the moments (average value and variance) of the size of fluctuations are obtained. A more complex and interesting task is to obtain the first-passage time for process of upper bounds on the moments and first-passage time (*FPT*) of the observable *A*, which are obtained by calculating the moments and *FPT* of the process of upper bound. Characteristic functions, average values and variances of the first-passage time of reaching fluctuations of observables of the trajectory of the Markov chain of positive and negative levels are also obtained. Some general issues of the relationship between the theory of random processes (using the example of the risk theory used) and thermodynamics of trajectories are also considered.

## 1. Introduction

Physics strives to establish general laws that are valid for wide classes of systems, ideally universal ones. One of these general, but perhaps not universal, patterns was obtained in Ref. [1], where proved the existence of general upper bounds on the size of fluctuations of any linear combination of fluxes (including all time-integrated currents or dynamical activities) for continuous-time Markov chains. These "inverse thermodynamic uncertainty relations" (i*TUR*s) are valid for all times and not only in the long-time limit. In paper [2], a generalization to first-passage times (*FPT*) of the so-called "inverse thermodynamic uncertainty relations" is carried out, which place an upper limit on the size of fluctuations in quantities integrated over time. The paper [2] specifically examines the *FPT* of counting observables, that is, the time of reaching a certain threshold of a trajectory value that takes values in positive integers and does not decrease with time. In [2], these upper bounds (or i*TUR*s) are extended to *FPT* counts of observable quantities, i.e. quantities corresponding to the total number of changes in the state of the system over time.

The lower limits of the size of fluctuations were obtained from thermodynamic uncertainty relations (*TUR*). Thermodynamic uncertainty relations Refs. [3-8] has significant generality. They turned out to be successful in describing various phenomena: the variance of time-averaged currents in the stationary state of continuous-time Markov chains, dynamics and observables, including finite times, discrete-time Markov dynamics, first-passage times, and open quantum systems, and many other extensions.

In Ref. [1], upper bounds for the fluctuation of observed trajectories were introduced, consisting of a linear combination of continuous-time Markov chain flows, including all flows and activities. The authors of Ref. [1] call them inverse *TUR*s (*iTUR*s). They are valid at all times and limit fluctuations at all levels. The resulting relations take into account the increase in fluctuations when approaching dynamic phase transitions between metastable states, which are not taken into account in the *TUR*.

In Ref. [1], the theory of large deviations (*LD*) is used, Ref. [9]. In Ref. [9], theorems were proven that, at large times, establish a correspondence between random variables and average values. For example, strong law of large numbers is true (Section 2.1). In Ref. [1], estimates were carried out not for random variables, but for dispersion, probabilities of deviation from average values, relative errors (the ratio of dispersion to the square of the average value).

These results are based on the expressions proven in Ref. [1] for the moment generating function of the form Eq. (4) for the quantity *A(t)* Eq. (1). The process *A(t)* characterizes the size of fluctuations. In work [1], it is shown that the moment-generating function (*MGF*) for the process *A(t)* is bounded from above and an expression for the bounding *MGF* is obtained. In this paper, *FPT* is defined for a process associated with a limiting *MGF*. The definition of *MGF* is given in subsection 2.3. From the results of Ref. [1], only expression (4) is used below. Knowing this value makes it possible to find not only the results mentioned above, but also many other important relationships: expressions for the moments and *FPT*, Refs [10-36], of the observable *A*, which are obtained by calculating the moments and *FPT* of the accumulated process. Determining *FPT* for process *A* (1) with cumulant (4) likely has many interesting and important applications, but faces difficulties. For example, *FPT*s for cumulant (4) to achieve negative levels have not been previously obtained in statistical physics. In this paper, we obtain the characteristics of the *FPT* for the upper-bound moment generating function of *A(t)* of the form (4). Boundary values of means and variances were obtained. For example, in (37) the values of mean *FPT* $\overline{\tau}_*(\gamma)$ are less others $\overline{\tau}(\gamma)$ because it corresponds to the cumulant $\tilde{\Lambda}$ (4) with maximum fluctuations when $\overline{\tau}(\gamma)$ are minimum. Let us note the importance of *FPT* in statistical physics (and not only in it).

Of significant interest is the explicit form of the boundary moments of the random variables of their size and their fluctuations, and their *FPT*. In this article these expressions are derived. It is assumed that, using the moment generating function corresponding to the upper limit of fluctuations, it is possible to obtain expressions for moments that also take on boundary values. While obtaining expressions for the size of the fluxes and their fluctuations is a relatively simple task, obtaining expressions for *FPT* moments is more difficult. The results of Refs. [37] and [38] are used. The results of risk theory were used. In this case, close connections were discovered between the relationships of the theory of random processes and the thermodynamics of trajectories. Some general issues of the relationship between the theory of random processes (using the example of the risk theory used) and statistical physics and thermodynamics of trajectories are also considered. As in Ref. [37], for a two-level model it is possible to consider the dependences of the moments on the total change in entropy in the system.

The level of achievement by a random process that reach the *FPT* can be arbitrary and exceed the average values of upper bounds on the size of fluctuations. In Ref. [37], in addition to expressions for the average values and variances of the quantity *A* and the *FPT* associated with it for the thermodynamics of trajectories (note that the results of Ref. [1] are more general and relate not only to the thermodynamics of trajectories), expressions were also obtained for the correlation between *A* and *FPT*. These expressions, obtained from the moment generating function of *A* of the form (4), will be boundary. Obtaining expressions for characteristic functions, average values and variances of the *FPT* value from expression (4) is associated with difficulties associated with solving equation (11). This paper uses the results of Ref. [38], where the average *FPT* values are obtained to achieve positive and negative bounds.

The article is organized as follows. In Section 2 upper bounding the moment generating function of *A* from Ref. [1] recorded and obtaining a simple correspondence between the relations of thermodynamics of trajectories and the theory of random processes for *FPT*. Relations for characteristic functions of *FPT* are obtained in Section 3. Explicit expressions for the moments of random variables of dynamic activity and *FPT* are obtained in Section 4. Brief conclusions are presented in Section 5.

Let us note the main results of this article:

**Main results**

- Explicit expressions for the upper bounds on the size of fluctuations of fluxes and for the moments of *FPT* fluctuations of observables of the trajectory ***X*** are obtained. Section 4 contains calculation formulas and graphs of calculated dependencies for the model of Section 2.1. Expressions for the moments of *FPT* are obtained in the Ref. [2], but in a general form and to *FPT* counts of observables, i.e., quantities corresponding to the cumulative number of changes in the state of the system over time. In the Ref. [2], the times of reaching a certain threshold value of the trajectory, which takes values in positive integers and does not decrease with time, are considered.
- The Ref. [2] considers the achievement of the number of jumps of the stochastic process. This value is non-negative. In this article, expressions are also obtained for achieving negative levels. The example in section 2.1 below contains negative values of the observables of the trajectory ***X***. The results of calculations to achieve negative boundary values are shown in Fig.6.
- The links between the theory of random processes and the thermodynamics of trajectories are established (see points 1)-11) in the Conclusion). Particular attention is paid to the problem of first-passage time for processes with independent increments in risk theory. We hope that such relations will be useful both in statistical physics and in the theory of random processes. An example of similar approaches in statistical physics and in the theory of random processes can be the non-equilibrium statistical operator method, Refs. [63, 64], in which the statistical operator is averaged over time with an exponential distribution, Ref. [62], and a randomly stopped process [61], in which the same averaging is performed, expression (10).

In table summarizes new results presented in this work.

| Section, formula | Results |
|---|---|
| Sec. 2, (18), (21), (23), (24) | Obtaining a simple correspondence for FPT between the relations of the theory of random processes and thermodynamics of trajectories |
| (29) | Probabilities $E(\tau(x)<\infty)$ are written explicitly (for example, Eq. (29)), which, as a rule, are not taken into account in statistical physics. |
| Sec. 4 | General expressions for average values and correspondence between risk theory and thermodynamics of trajectories |
| (37) | The inequalities for upper and lower bounds of means and variances |
| (38) | Obtaining a more accurate expression for *SCGF* $g(\mu)$ |
| (39), (40) | Accounting for the influence of risk theory |
| Sec. 4.1-4.3; (42)-(43) | Possibilities of generalizing the results thermodynamics of trajectories: consideration of cases $m<0$ and $x<0$ |
| (49)-(57) | The solution of the Lundberg equation Expressions for $\rho_+$, $\rho_-$, the means and variance of *FPT* achieving positive and negative level for the four-state model |
| Conclusion | Connections and mutual influence of the theory of random processes and statistical physics, risk theory and thermodynamics of trajectories |
| Conclusion | 5). After differentiation by $\gamma$, when determining the average values, it is not necessary to assume $\gamma=0$, since this corresponds to the equilibrium case, and in non-equilibrium situations $\gamma\neq 0$ |
| Conclusion | 9). The important role of risk theory in describing *FPT* processes with cumulant Eq. (4). |

## 2. Obtaining a simple correspondence between the relations of the theory of random processes and thermodynamics of trajectories for FPT

Having upper and lower bounds on the variation of a random variable or process is important because they restrict the range of values for fluctuations. The inverse *TUR* and *TUR* gives a range that limits the probability of observing a fluctuation from the mean (rather than just a one-sided bound). It is shown in Ref. [1] that *iTUR* can often be a tighter bound than *TUR*. This can provide an advantage for estimation, which is offset by the fact that although *iTUR* also depends on general system characteristics (relaxation time, shortest time scale), it is not as operationally accessible as *TUR*. A possible area for further study is the application of *iTUR* in constructing confidence intervals for parameter estimation, Ref. [2].

In Section 2 we write a bounding moment generating function for the process of the considered fluctuations of trajectory observables. Also, in this section the results of the study of boundary value problems, namely the problem of achieving *FPT* for processes with independent increments in risk theory, are compared with the results of statistical physics using the example of thermodynamics of trajectories. Some important relations of the theory of random processes have direct analogies with the expressions of thermodynamics of trajectories.

### 2.1. Definition of an upper bounding the moment generating function

This article discusses *FPT* for a cumulant process Eq. (4). Let us define this quantity. In Ref. [1], an explicit form of the cumulants of the considered fluctuations of trajectory observables was obtained. Consider a Markov chain $\boldsymbol{X}:=(X_t)_{t>0}$ with continuous time in a finite state space $E$ with a generator $W=\sum_{x\neq y} w_{xy}\left|x\right\rangle\left\langle y\right|-\sum_{x} w_{xx}\left|x\right\rangle\left\langle x\right|$, with $x,y\in E$. In work [1], general upper bounds on the size of fluctuations of any linear combination of fluxes, including all time-integrated currents or dynamic activities, were investigated. Let us assume that $X_0$ is distributed according to some probability measure $v$ in the state space. We denote by $P_v$ the law of $\boldsymbol{X}$ and we use $\boldsymbol{E}_v$ for the corresponding expected value. We assume that $\boldsymbol{X}$ is irreducible with unique invariant measure (i.e., stationary state) $\pi$. The subject of the study will be fluctuations of observables of the trajectory $\boldsymbol{X}$ of the form:

$$A(t)=\sum_{x\neq y} a_{xy} N_{xy}(t)\,. \tag{1}$$

In Eq. (1) $a_{xy}$ are arbitrary (in contrast with Ref. [7], where $\alpha_{xy}\geq 0$) real numbers with $\sum\left|a_{xy}\right|>0$ , and $N_{xy}(t)$ are the elementary fluxes, that is, the number of jumps from $x$ to $y$ up to time $t$ in $\boldsymbol{X}$. For a time-integrated current the coefficients are antisymmetric, $a_{xy}=-a_{yx}$, while for counting observables (such as the activity), they are symmetric, $a_{xy}=a_{yx}$. Therefore, size of fluctuations of any linear combination of fluxes (including all time-integrated currents or dynamical activities) can be considered; $A(t)$ Eq. (1) is equal to activity $K$ when given in Eq. (1) $a_{xy}=1$. In Ref. [1], it is shown that for every $u\geq 0$ the moment generating function (*MGF*) of $A(t)$ is:

$$Z_{\pi,t}(u)=\boldsymbol{E}_{\pi}[e^{uA(t)}]=\left\langle \pi\left|e^{tW_u}\right|-\right\rangle, \tag{2}$$

where $|-\rangle = \sum_x |x\rangle$. In the stationary state *π*, the average of *A* per unit time in the long time is $\langle a \rangle_\pi = \sum_{x\neq y} \pi_x w_{xy} a_{xy}$ ($A(t) \approx t\langle a \rangle_\pi$, strong law of large numbers), while its static approximate variance is $\langle a^2 \rangle_\pi$, with $\langle a^2 \rangle_\pi = \sum_{x\neq y} \pi_x w_{xy} a^2{}_{xy}$ (it is the variance of the random variable $\sum_{x\neq y} a_{xy} \tilde{N}_{xy}$, where we approximate $N_{xy}(t)/t$ with independent Poisson random variables $\tilde{N}_{xy}$ with intensity $\pi_x w_{xy}$). The maximum escape rate is $q=\max_x w_{xx}$, and $c = \max_{x\neq y} |a_{xy}|$ is the maximum amplitude of the coefficients that define the observable. Since we do not assume that *W* is reversible, we denote by *ε* the spectral gap of the symmetrization $R(W)=(W+W^+)/2$, where the adjoint is taken with respect to the inner product induced by the stationary state *π*. The average dynamical activity per unit of time at stationarity is $\langle k \rangle_\pi = \sum_{x\neq y} \pi_x w_{xy}$; the "tilted" generator $W_u = \sum_{x\neq y} e^{ua_{xy}} w_{xy} |x\rangle\langle y| - \sum_x w_{xx} |x\rangle\langle x|$, which is an analytic perturbation of *W*. In work [1], the titled generator is defined as $W_u = \sum_{x\neq y} (e^{ua_{xy}} - 1) w_{xy} |x\rangle\langle y| + W$. At long times:

$$Z_{v,t} \le C(v) e^{t\tilde{\Lambda}(u)}, \tag{3}$$

where $C(v) = (\sum_x v_x{}^2 / \pi_x)^{1/2}$ accounts for the difference between *v* and the stationary *π*, with $C(\pi)=1$. At long times $Z_{\pi,t} = \boldsymbol{E}_\pi[e^{uA(t)}] \sim e^{t\Lambda(u)}$, where the scaled cumulant generating function (*SCGF*) (or cumulant for short, we will use both terms) $\Lambda(u)$ is the largest eigenvalue of $W_u$. In work [1], for *SCGF* $\tilde{\Lambda}(u)$ from Eq. (3), the following expression was obtained:

$$\tilde{\Lambda}(u) = \sum_{x\neq y} \pi_x w_{xy} (e^{ua_{xy}} - 1) + \frac{q\langle a^2 \rangle_\pi u^2}{\varepsilon(1 - 5qcu/\varepsilon)}, \tag{4}$$

if $0 \le u < \varepsilon / 5qc$ and $+\infty$ otherwise. In work [1], it is shown that $2q/\varepsilon \ge 1$. In works [39] and [40], the fundamental role of the spectral gap of the system in characterizing its dynamics is noted.

Main results in Ref. [1] for this article are for the moment generating function (*MGF*) Eq. (3) and *SCGF* Eq. (4) of *A(t)*. Based on these expressions, which are of independent interest, the results indicated in Ref. [1] were obtained. Estimate Eq. (3) consists of two parts in expression (4), where the first summation in Eq. (4) is the *SCGF* of $\sum_{x\neq y} a_{xy} \tilde{N}_{xy}$, where we approximate $N_{xy}(t)/t$ with independent Poisson random variables $\tilde{N}_{xy}$ with rates $\pi_x w_{xy}$; the second term takes into account correlations between jumps in the Markov chain. The basis for the results obtained in Ref. [1] are the expressions for *SCGF* $\tilde{\Lambda}(u)$ Eq. (4) calculated by summation in Ref. [1]. This *SCGF* is the upper limit for all *SCGF*s of process Eq. (1) in this sense provides universal expressions for the process under consideration of the form Eq. (1).

From the *SCGF* of the form Eq. (4) we obtain the average values of *A(t)*. If we determine the *SCGF* of the form *g(γ)* Eq. (7), then we can find the average values of the *FPT*. Of interest are the dependences of such quantities on the parameter *γ*. Dependencies of this kind were obtained in Ref. [37], and the parameter *γ*, conjugate to the *FPT* Ref. [35], is expressed in terms of the change in entropy.

### 2.2. The thermodynamics of trajectories, s- and *x*-ensembles

It was shown in Ref. [38] that an explicit expression for the first passage time from a *SCGF* of the form Eq. (4) can be obtained using risk theory, Refs. [57-59].

Below, the obtained results and the used relations are compared with the thermodynamics of trajectories, Refs. [41-56]. Therefore, we will show how the so-called $s$-ensemble and $x$-ensemble are defined in the thermodynamics of trajectories. The partition function (or the moment generating function) from relations Eqs. (2)-(3), for example, in the thermodynamics of trajectories, Refs. [41-56], (below are examples from this area) is equal to:

$$Z_\tau(s) \equiv \sum_K e^{-sK} P_\tau(K) = \sum_{\boldsymbol{X}_\tau} e^{-s\hat{K}[\boldsymbol{X}_\tau]} P(\boldsymbol{X}_\tau) \sim e^{\tau\theta(s)}, \qquad \frac{\langle\langle K^n \rangle\rangle}{\tau} = (-1)^n \frac{\partial^n}{\partial s^n}\theta(s)_{|s=0}, \tag{5}$$

where $P_\tau(K)$ is distribution of value $K$ over all trajectories $\boldsymbol{X}_\tau$ of total time $\tau$, $\mathbf{X}_\tau = (C_0 \to C_{t_1} \to ... \to C_{t_K})$, $P(\mathbf{X}_\tau)$ is the probability for observing this trajectory out of all the possible ones of total time $\tau$; $\langle\langle . \rangle\rangle$ indicates cumulant (mean, variance, etc.). The value $K$ defined as the total number of configuration changes in a trajectory. This trajectory has $K$ jumps, with the jump between configurations $C_{t_{i-1}}$ and $C_{t_i}$ occurring at time $t_i$, with $0 \le t_1 \le \cdot\cdot\cdot t_K \le t$, and no jump between $t_K$ and $\tau$. We within the ensemble of trajectories of total time $\tau$. For $x$-ensemble Eqs. (6)-(7) is kept fixed is the total number of configuration changes $K$ on trajectories $Y_K = (C_0 \to C_{t_1} \to ... \to C_\tau)$, where the number of configuration changes is fixed at $K$, but the time $\tau$ of the final $K$-th jump fluctuates from trajectory to trajectory. The proportionality $Z_\tau(s)$ of the function $e^{\tau\theta(s)}$ in Eq. (5) is obtained using the theory of large deviations (*LD*).

Expressions for *SCGF*s Eqs. (3)-(4) refers to the so-called $s$-ensembles with *MGF* Eq. (5), Refs. [41-49, 1, 8, 50, 55], of the thermodynamics of trajectories in which the process time $\tau$ is fixed, and the dynamic activity $K$, the number of events, changes in the trajectory during time $\tau$ is a random variable. In Refs. [49, 56], the so-called $x$-ensembles are considered, in which the values of dynamic activity $K$ are fixed, and the time to reach a fixed value of $K$, *FPT* is a random variable. Distributions in which *FPT* (lifetime) is a thermodynamic variable were introduced in Refs. [13, 34-35]. The conjugate *FPT* thermodynamic parameter $\gamma$ (equal to parameter $x$ from Refs. [56, 49], just another designation) was associated with entropy changes in the system. Let us replace the notation $x$ with $\gamma$, since below $x$ denotes the level of achievement by a random process. The notation $\gamma$ was used in Refs. [13, 34-35], the notation $x$ in works [56, 49]. In Refs. [13, 34-35], the parameter $\gamma$ is conjugate to the random *FPT* in the distribution.

The partition function, the corresponding moment generating function for random time, of reaching a fixed value $K$ of dynamic activity is (in $x$-ensemble, at $x$=$\gamma$):

$$Z_K(\gamma) = \int_0^\infty d\tau e^{-\gamma\tau} P_K(\tau), \tag{6}$$

where $P_K(\tau)$ is the distribution of total trajectory length $\tau$ for fixed activity $K$. In Ref. [35], the quantity corresponding to $Z_K(\gamma)$, the Laplace transform of the *FPT* distribution, acts as a non-equilibrium partition function. For large $K$ the generating function also has a large deviation (*LD*) form:

$$P_K(\tau) \sim e^{-K\Phi(\tau/K)}, \qquad Z_K(\gamma) \sim e^{Kg(\gamma)}. \tag{7}$$

The total fixed number of configuration changes, i.e., the dynamical activity $K$ related with the average dynamical activity per unit time at stationarity $\langle k \rangle_\pi = \sum_{x \ne y} \pi_x w_{xy}$ from Eq. (5) and Section 2.1, $K \approx t \langle k \rangle_\pi$. When $K$, $\tau$ are large, $K$, $\tau >> 1$, relations (5), (7) have large deviation form. The function $g$ is the functional inverse of $\theta$ and vice versa (Refs. [47, 57]):

$$\theta(s) = g^{-1}(s), \quad g(\gamma) = \theta^{-1}(\gamma), \quad s = g(\gamma), \quad \gamma = \theta(s)\,. \tag{8}$$

### 2.3. Definitions of the characteristic functions and Lundberg equation

Let us define the quantities that are used below. The characteristic function of a homogeneous process $\xi(t),\ t \ge 0$ is determined in the theory of random processes, Refs. [60, 57], by the relation (if $\xi(0) \ne 0$):

$$Ee^{i\alpha(\xi(t)-\xi(0))} := \int_{-\infty}^{\infty} e^{i\alpha x} dF(x) = e^{t\Psi(\alpha)},\ \ t \ge 0\,, \tag{9}$$

where $F(x) = P(\xi < x)$ is distribution function of a random process $\xi(t),\ t \ge 0$, the function $\Psi(\alpha)$ is the cumulant (or *SCGF*) of the process $\xi(t),\ t \ge 0$. The characteristic function Eq. (9) corresponds to the partition function (or *MGF*) Eq. (5) (for *is*=*α*, *iα*=-*s*), and the function $\Psi(\alpha)$, the cumulant of the process $\xi(t),\ t \ge 0$ corresponds to the cumulant *θ(s)* Eq. (5).

Let $X$ be a random variable with cumulative distribution function $F_X$. The moment-generating function (*MGF*) of $X$ (or $F_X$), denoted by $M_X(t)$, is $M_X(t)=\boldsymbol{E}[e^{tX}]$. If $X$ is a continuous random variable, the following relation between its moment-generating function (*MGF*) $M_X(t)$ and the two-sided Laplace transform of its probability density function $f_X(x)$ holds: $M_X(t) = L\{f_X\}(-t)$, since the probability density function two-sided Laplace transform is given as $L\{f_X\}(s) = \int_{-\infty}^{\infty} e^{-sx} f_X(x)dx$ and the moment-generating function's definition expands (by the law of the unconscious statistician) to $M_X(t) = E[e^{tX}] = \int_{-\infty}^{\infty} e^{tx} f_X(x)dx$.

For the characteristic function of the process $\xi(\theta_s)$ defined as the Laplace–Carson transform $Ee^{i\alpha\xi(\theta_s)} = s\int_{-\infty}^{\infty} Ee^{i\alpha\xi(t)} e^{-st} dt$, the relation is written:

$$\varphi(s,\alpha) := Ee^{i\alpha\xi(\theta_s)} = \frac{s}{s-\Psi(\alpha)}\,. \tag{10}$$

The characteristic function Eq. (10) is the Laplace–Carson transforms of the characteristic function Eq. (9) (at $\xi(0)=0$). We consider $\xi(\theta_s)$ as a randomly stopped process. Randomly stopped processes, Ref. [61], contain an exponentially distributed random variable $\theta_s$ independent of the process $\xi(t)$, $P\{\theta_s > t\} = e^{-st},\ \ s > 0,\ \ t > 0$. The same procedure, as shown in Ref. [62], is carried out in the non-equilibrium statistical operator method works [63, 64].

If for a process $\xi(t),\ t \ge 0$ the denominator in Eq. (10) at $i\alpha = r$ is equal to zero, then we obtain the equation:

$$\Psi(\alpha)\big|_{i\alpha=r} := k(r) = s, \quad \pm \mathrm{Re}\, r \ge 0\,, \tag{11}$$

which in risk theory, Refs. [57, 58, 59], is called the fundamental Lundberg equation. Note that for the cumulant, the designations $\Psi(\alpha)$ are used, as in Eq. (10), and the designations $\Psi(\alpha)\big|_{i\alpha=r} = k(r)$ are used in Eq. (11). When replacing *iα*=*r*, *ir*=-*α*, in Eq. (9) $\boldsymbol{E}e^{i\alpha\xi(t)}$ is replaced by $\boldsymbol{E}e^{r\xi(t)}$ as in *MGF*. In Ref. [65], the function $\Psi(-i\alpha)$ was replaced by a function $\Psi_1(\alpha)$ corresponding to *k(r)*. This function will be convex on a segment $(\infty\ ,\ \alpha_+]$ of the real axis $(0 \le \alpha_+ = \sup\{\alpha : k(\alpha) < \infty\})$, $k'(0) = E\xi(1) = a,\ \ k''(0) = D\xi(1) > 0$. If the second derivative of a function is positive, the function is convex. Equation (11) plays an important role in subsequent calculations. The equation (11) has two roots; $r_1 = -\rho_-(s),\ r_2(s) = \rho_+(s) > 0$. The solution to equation (11) is given in Section 4.6 using the example of a four-state system.

The function $\psi(\alpha)$ from Eq. (9) is the *SCGF* of the process $\xi(t),\ t \geq 0$, the Levi-Khinchin form is used in Eq. (11), works [57, 65]; for a homogeneous process with independent increments $\{\xi(t),\ t \geq 0, \xi(0) = 0\}$ of the cumulant is:

$$\Psi(\alpha) = i\alpha\gamma - \frac{1}{2}\alpha^2\sigma^2 + \int_{-\infty}^{\infty}(e^{i\alpha x} - 1 - \frac{i\alpha x}{1+x^2})\Pi(dx), \quad \int_{|x|\leq 1} x^2\Pi(dx) < +\infty,\ \sigma^2 \geq 0,$$

*γ* has an arbitrary sign. The convexity of a function $\Psi(\alpha)$ can be directly verified from this formula. Due to the convexity $k(r)$ in the neighborhood of zero ($k``(0) > 0$) for sufficiently small *s* for continuous from above (from below) $\xi(t), t \geq 0$, equation (11) has a positive root $r_s = \rho_+(s) > 0$ (negative root $r_s = -\rho_-(s) < 0$). The value *m* is defined as:

$$m = \partial k(r)/\partial r\big|_{r=0} = \langle a \rangle_\pi, \tag{12}$$

where $\langle a \rangle_\pi$ from Section 2.1. The behavior of the roots $\rho_\pm(s)$ at $s \to 0$ depends on the mean value $m = \boldsymbol{E}\xi(1), |m| < \infty, D\xi(1)$, *D* is the variance.

It was shown in Ref. [57] that for $m = k'(0) = 0,\ \rho_\pm(s) \approx \sqrt{2s/D\xi(1)}$, and for $m > 0,\ (m < 0)$ $\rho_+(s) \approx m^{-1}s \quad (\rho_-(s) \approx |m|^{-1} s,\ s \to 0)$. If $\pm m$<0, then $\rho_\pm(s)_{s\to 0} \to \rho_\pm > 0$.

If a process with positive jumps is monotonically nondecreasing, then Eq. (11) has no negative roots. If a process with negative jumps is monotonically non-increasing, then equation (11) with $s > 0$ has no positive roots.

Let's return to the *s*-ensemble Eq. (5). In work [8], lower bound for the scaled cumulant generating function *θ(s)* for process Eq. (1) with positive values $a_{xy}>0$ is written in a form different from Eq. (4):

$$\theta(s) \geq \theta_*(s) = \langle k \rangle (e^{-s\langle a \rangle_{\pi+}/\langle k \rangle} - 1), \tag{13}$$

where $\langle k \rangle = \langle K \rangle / \tau = \sum_{x\neq y} \pi_x w_{xy} = \langle k \rangle_\pi$ (Section 2.1) is the average dynamical activity (per unit time). Inverting $\theta_*(s)$ in Eq (13) provide a lower *SCGF* *g(μ)* (from Eqs. (6)-(8)):

$$g(\mu) \geq g_*(\mu) = -\frac{\langle k \rangle}{\langle a \rangle_{\pi+}} \ln(1 + \frac{\mu}{\langle k \rangle}), \tag{14}$$

The Laplace transformed *FPT* distribution $P_K(\tau)$ in Eqs. (6), (7) has a large deviation form $Z_K(\gamma)$ in Eq. (7), where $P_K(\tau)$ is the *FPT* distribution; *τ* is the *FPT* through *A(ω)=A*; *FPT* are events for a fixed value *A* of the observable *A(ω)*, the state in the finite state space *E*, and ending after *A* jumps [8], *A(ω)* is observable in trajectory *ω*. The *SCGF* $\theta_*(s)$ in Eq. (13) is close to the first term in Eq. (4), but differs from it in that the $a_{xy}$ values in Eq. (4) are arbitrary real numbers from Eq. (1), and expression (13) in work [8] was obtained for the process in which these quantities are positive.

### 2.4. *x*-ensemble and cumulant for FPT; a correspondence between the quantities of the theory of random processes and the thermodynamics of trajectories

In this subsection, a correspondence is established between *MGF* and *SCGF* of *x*-ensembles of the thermodynamics of trajectories and cumulants of the random *FPT* process. We use the definitions for $\tau^{+}(x)$, Ref. [57], $\tau^{+}(x)=\inf\{t:\xi(t)>x\},\ x>0$, is the moment of the first exit of the process $\xi(t)$ for the level $x>0$.

From expressions (8), (13) we obtain the expression (14), at $g(\gamma)\approx g_{*}(\gamma)$. From here and from Eqs. (6)-(7), (14), we obtain:

$$\langle\tau\rangle_{\gamma}=-\frac{\partial\ln Z_{K}(\gamma)}{\partial\gamma}=-K\frac{\partial g(\gamma)}{\partial\gamma}=\frac{K}{\langle a\rangle(1+\gamma/\langle k\rangle)},\quad K=\langle K\rangle=\langle K_{\gamma=0}\rangle, \tag{15}$$

where $\langle K_{\gamma}\rangle=\sum_{K}Ke^{-sK}P_{\tau}(K)/Z_{\tau}(s)\big|_{s=g(\gamma)}$ - ensemble average with partition function (5) using Eqs. (8). Since Eqs. (8) are obtained for $K,\ \tau$ are large, $K,\ \tau>>1$, the same limitations apply to expression (15). The last equality in Eq. (15) is written from the results of Ref. [37].

Let us compare the cumulant:

$$\Psi(\alpha)=\int_{0}^{\infty}(e^{i\alpha x}-1)\Pi(dx) \tag{16}$$

(for $is=\alpha$) with expression (13) ($\Pi(dx)$ is probability measure proportional to the probability distribution, satisfying the conditions specified in work [60]). We use the expression for the *MGF*, Refs. [38, 57]:

$$T(s,x)=E[e^{-s\tau^{+}(x)},\ \tau^{+}(x)<\infty]=P\{\xi^{+}(\theta_{s})>x\}=e^{-\rho_{+}(s)x},\ x\geq 0\,, \tag{17}$$

where $\xi^{+}(\theta_{s})$ is the maximum of the process $\xi(\theta_{s})$. Note that this is a relatively simple type of generatrix. On p. 90 in Ref. [57], Lemma 3.1 is proved: For an upper-continuous process $\xi(t)$ with cumulant Eq. (11) or Eq. (13), the *MGF* $T(s,x)$ is determined by the relation:

$$T(s,x)=e^{-\rho_{+}(s)x},\quad \rho_{+}(s)=k^{-1}(s),\quad x\geq 0\,. \tag{18}$$

The index $\rho_{+}(s)$ is the positive root of the cumulant equation Eq. (11) of the form $\Psi(-i\alpha)=k(r)=s$.

In expression (16) we substitute:

$$\Pi(dx)=\langle k\rangle\delta[x-\langle a\rangle_{\pi+}/\langle k\rangle]. \tag{19}$$

Then for $i\alpha=r$ (as in Eq. (11)) we obtain:

$$\Psi(\alpha)_{|\alpha=r}=k(r)=\langle k\rangle(e^{r\langle a\rangle_{\pi+}/\langle k\rangle}-1). \tag{20}$$

Substitution $r=-s$ in Eq. (20) leads to equality of Eq. (20) and Eq. (13) and to the relation:

$$\theta_{*}(s)=k(r=-s). \tag{21}$$

If we put in the Lundberg equation Eq. (11) $s=\gamma$, then we will find from the equation $k(r)=\gamma=\langle k\rangle(e^{r\langle a\rangle_{\pi+}/\langle k\rangle}-1)$, assuming $r=r_{s}=\rho_{+}(\gamma)$ that:

$$\rho_{+}(\gamma)=[\langle k\rangle/\langle a\rangle_{\pi+}]\ln[1+\gamma/\langle k\rangle]. \tag{22}$$

Comparing expressions (22) and Eq. (14) for $\mu=\gamma$, we obtain:

$$\rho_{+}(\gamma)=-g_{*}(\gamma). \tag{23}$$

Correspondence (at $g_{*}(\gamma)\rightarrow g(\gamma)$) with expression (8) is established. Thus, the expression $s=g(\gamma)$ corresponds to expression (21), since $s=g(\gamma)=-\rho_{+}(\gamma)=-r_{s}=s$. The expression $\gamma=\theta(s)$ from Eq. (8) corresponds to equation (14) with $s=\gamma$ and relation (21) (at $\theta_{*}(s)\rightarrow\theta(s)$). The expression $\theta(s)=g^{-1}(s)$ from Eq. (8) corresponds to expressions (18), (21), (23). The

expression $g(\gamma)=\theta^{-1}(\gamma)$ from Eq. (8) corresponds to expressions (18), (21), (23). The boundary value $K$ from expressions (6)-(7), (15) corresponds to the value $x$ from Eqs. (17)-(18), Section 4 and expressions (25), (29), (31)-(33), (38)-(43), (48)-(49).

Thus, the solutions to the problem of determining the random variable *FPT* $\tau^+(x)$ obtained in the theory of random processes Eqs. (16)-(23) and in the thermodynamics of trajectories Eqs. (14)-(15) coincide under conditions Eqs. (18)-(23). This happens because the second expression (18) of the theory of random processes coincides with the relations (8) of the thermodynamics of trajectories.

Thus, a correspondence of the form is established between the quantities of the theory of random processes and the thermodynamics of trajectories:

$$K\to x,\quad g(\gamma)\to-\rho_+(\gamma),\quad E[e^{-\gamma\tau^+(x)},\tau^+(x)<\infty]\to Z_K(\gamma)\sim e^{Kg(\gamma)}. \quad (24)$$

However, for process Eq. (1) with *SCGF* Eq. (4), relations (24) are generalized (for example, expressions (31), (38)). The characteristic functions and *SCGF*s of real physical processes apparently have a more complex form than Eqs. (13), (14). Another important generalization associated with risk theory Ref. [57] is the ability to describe the achievement of negative values, as opposed to the achievement of only positive values, in thermodynamics of trajectories.

## 3. Relations for characteristic functions of FPT

In Ref. [57], results for characteristic functions corresponding to cumulants of a certain type were proven. In order to use the results of Ref. [57], it is necessary to reduce *SCGF* Eq. (4) to this form. This operation is done in section 3.

### 3.1. Characteristic function for FPT reaching a positive level *x*>0

In Ref. [57], the ratio is established for the pair $\{\tau^+(x),\gamma^+(x)\}$:

$$\boldsymbol{E}[e^{-s\tau^+(x)-u\gamma^+(x)},\tau^+(x)<\infty]=\boldsymbol{E}[e^{-u\gamma^+(x)},\xi^+(\theta_s)>x]=\boldsymbol{E}e^{-u\gamma^+(x)}P\{\xi^+(\theta_s)>x\}=\frac{c_1}{c_1+u}q_+(s)e^{-\rho_+(s)x},\ x>0,\ (25)$$

where $x$>0, $\gamma^+(x)=\xi(\tau^+(x))-x$ is first overjump over $x$>0; $\tau^+(x)=\inf\{t:\xi(t)>x\}$, $x>0$ is the moment of the first exit of the process $\xi(t)$ for the level $x$>0; $\zeta^{\pm}(t)=\sup_{0\le t'\le t}(\inf)\zeta(t')$ represents the extremes of the process $\zeta(t')$ on the interval [0, $t$]; $c_1$(=$\varepsilon/5qc$); at $m$<0, $r_s=\rho_+(s)=c_1p_+(s)<c_1$, $\rho_+(s)_{s\to0}\to\rho_+=c_1p_+>0$, $q_+(s)=1-p_+(s)$. In Ref. [38], it is shown (see also Appendix A, (A.1), (A.4)) that the *SCGF* of the form Eq. (4) with the approximations described in Appendix A corresponds to the *SCGF* of the form:

$$k(r)=-\frac{\alpha_-\bar{k}r}{m_{3-}+r}+\frac{\alpha_+\bar{k}r}{m_{3+}-r}+\frac{\lambda_1c_1r}{c_1-r}-\lambda_1r,\ \ m_{3\pm}=\langle k\rangle/\langle a\rangle_{\pi\pm},\ \ \langle a\rangle_{\pi-}=\langle a\rangle_\pi\big|_{a_{xy}<0},\ \langle a\rangle_{\pi+}=\langle a\rangle_\pi\big|_{a_{xy}>0},\ (26)$$

$$\lambda_1=\frac{\langle a^2\rangle_\pi}{5c},\quad c_1=\frac{\varepsilon}{5cq},\quad \bar{k}=\sum\nolimits_{x\ne y}\pi_xw_{xy}=\langle k\rangle_\pi,\quad f_1=\frac{\pi_xw_{xy}}{\bar{k}},\quad \frac{\partial k(r)}{\partial r}\Big|_{r=0}=m=\langle a\rangle_\pi=\sum\nolimits_{x\ne y}\pi_xw_{xy}a_{xy}.$$

In expressions (26) an exponential distribution model Eq. (A.2) is used for the distribution density of the form $\pi_xw_{xy}/\bar{k}$, $\bar{k}=\sum\nolimits_{x\ne y}\pi_xw_{xy}$ (Appendix A, Eq. (A.2)), which is introduced in the first term of the *SCGF* Eq. (4). Such distributions are widely used in risk theory works [57-59]. However, it will be shown below that there are approximations that more accurately describe the behavior of *SCGF*s Eq. (4). Section 4 uses other approximations. In Ref. [38], the exponential distribution and cumulant Eq. (26) was used to compare *SCGF*s Eq. (4) with cumulants for which theorems were proven in Ref. [57], containing relations for the times of reaching positive and negative levels. In Appendix A, *SCGF* Eq. (4) is written in the form Eq. (A.1). This expression in Ref. [38] is reduced to the form for which relation (25) was proven in Ref. [57].

On condition *m*<0, $r_s = \rho_+(s) = c_1 p_+(s) < c_1$, $\rho_+(s)_{s\to 0} \to \rho_+ = c_1 p_+ > 0$, $q_+(s) = 1 - p_+(s)$. This condition is satisfied by the expression for $r_{s0} = r(s=0) = \rho_+$, $\rho_+$ is a positive root, solution of the equation:

$$k(r_s) = s\big|_{s=0}\,, \tag{27}$$

when *s*=0 is assumed in equation (11).

Equation (27) for cumulant Eq. (26) has three roots: zero, negative, and positive. The latter looks like:

$$\rho_+ = (\sqrt{b^2 - 4a_5 c_2} - b)/2a_1\,, \tag{28}$$

where $a_5 = b_2 \ge 0$ $\alpha_+ \le 0.35$, $b = \frac{\bar{k}}{|\bar{a}_-|}(b_2 + \alpha_+ \bar{a}_+) - \frac{\bar{k}}{|\bar{a}_-|}\bar{a} + c_1 \alpha_+ \bar{a}_+ > 0$, $\bar{a} < 0$, $c_2 = \frac{c_1 \bar{k}\bar{a}}{|\bar{a}_-|} < 0$, $\bar{a}_+ = \langle a \rangle_{\pi+} > 0$, $\bar{a} = \langle a \rangle_{\pi}$, $\bar{a}_- = \langle a \rangle_{\pi-} < 0$, $b_2 = \lambda_1 + \alpha_+[\langle a^2 \rangle_{\pi+}/2 - \langle a \rangle_{\pi+}]$. From Eqs. (25) and (28) we get:

$$\boldsymbol{E}[e^{-s\tau^+(x) - u\gamma^+(x)}, \tau^+(x) < \infty]\big|_{u=0,s=0} = \boldsymbol{E}[\tau^+(x) < \infty] = (1 - \frac{\rho_+}{c_1})e^{-\rho_+ x},\ x > 0\,. \tag{29}$$

### 3.2. Characteristic function for FPT reaching negative level *x*<0

In Ref. [57], results for the characteristic functions of reaching negative levels were obtained for a lower semi-continuous process with a cumulant of the form:

$$k(r) = -\frac{\lambda_2 r}{b_1 + r} + a_3 r + \int_0^\infty (e^{rx} - 1)\Pi(dx),\ \ a_3 \ge 0,\ \ b_1 > 0\,. \tag{30}$$

In Ref. [38], this cumulant Eq. (30) is compared with cumulants Eqs. (A.1), (A.4) identical to Eq. (4). More details about cumulant compliance are written in Ref. [38].

From the Lundberg equation (11) we obtain: $\frac{\partial k(r)}{\partial s} = 1 = \frac{\partial k(r)}{\partial r}\frac{\partial r}{\partial s}$, $\frac{\partial r}{\partial s} = 1/\frac{\partial k(r)}{\partial r}$, $\frac{\partial^2 r}{\partial s^2} = -\frac{\partial^2 k(r)}{\partial r^2}(\frac{\partial k(r)}{\partial r})^{-3}$. The expansion of a function $\rho_-(s)$ in powers up *s* to a quadratic term has the form $\rho_-(s) = \rho_-(s=0) + s(\partial\rho_-(s)/\partial s)_{s=0} + (s^2/2)(\partial^2\rho_-(s)/\partial s^2)_{s=0} + \ldots$. Substituting the above relations for $\partial r/\partial s_{s=0}$, $\partial^2 r/\partial s^2{}_{s=0}$, at $r_s(s) = -\rho_-(s)$, $r(s=0) = 0$, first solution to equation (27) $k(r)\big|_{s=0} = 0$, into this expression, we obtain from expressions (30), (A.1) at

$\bar{a} = \langle a \rangle_\pi = m < 0$, $\rho_-(s) = \frac{s}{|m|} - \frac{1}{2}s^2\frac{2}{|m|^3}[\frac{1}{\bar{k}}(\alpha_+\bar{a}_+{}^2 + \alpha_-|\bar{a}_-|^2) + \langle a^2 \rangle_\pi q/\varepsilon]$, $\bar{a}_+ = \langle a \rangle_\pi\big|_{a_{xy}>0}$,

$\bar{a}_- = \langle a \rangle_\pi\big|_{a_{xy}<0}$, $p_-(s) = \frac{1}{b_1}\rho_-(s)$, $r_s = -\rho_-(s) < 0$, $m < 0$, $\rho_-(s)_{s\to 0} \to 0$, $\rho'_-(0) = 1/|m|$, $q_- = 1 - p_-$, $r(s) = r_s = -\rho_-(s)$, $r_s(s=0) = 0$ is a negative root of the Lundberg equation (11).

Conditions $r < m_{3+}$, $r < c_1$ in Eqs. (A.2)-(A.4) are met. In work [57] for cumulant Eq. (30) the expression was obtained:

$$\boldsymbol{E}[e^{-s\tau^-(x)+u\gamma^-(x)}, \tau^-(x) < \infty] = \boldsymbol{E}[e^{u\gamma^-(x)}, \xi^-(\theta_s) < x] = \boldsymbol{E}e^{u\gamma^-(x)}P\{\xi^-(\theta_s) < x\} = \frac{c_1}{c_1+u}q_-(s)e^{\rho_-(s)x}, \; x<0 . \quad (31)$$

From Eq. (31) we obtain:

$$\left\langle \tau^-(x) \right\rangle = -\frac{\partial \boldsymbol{E}[e^{-s\tau^-(x)+u\gamma^-(x)}, \tau^-(x) < \infty]}{\partial s}\Big|_{u=0,s=0} = \frac{1}{|m|}(\frac{|\bar{a}_-|}{\bar{k}} - x), \;\; x<0 .$$

After inversion Eq. (31), the relation is written for $x$<0, $z$<0:

$$P\{\gamma^-(x) < z, \xi^-(\theta_s) < x\} = P\{\xi^-(\theta_s) < x\}P\{\gamma^-(x) < z\} = q_-(s)e^{\rho_-(s)x}e^{c_1 z} . \quad (32)$$

From Eq. (31) the expression for the variance is written:

$$D_{\tau^-} = \left\langle \tau^{-2}(x) \right\rangle - \left\langle \tau^-(x) \right\rangle^2 = \frac{1}{|m|^2}(\frac{|\bar{a}_-|}{\bar{k}} - x)\{\frac{2}{|m|}[\frac{1}{\bar{k}}(\alpha_+\bar{a}_+{}^2 + \alpha_-|\bar{a}_-|^2) + \left\langle a^2 \right\rangle_\pi q/\varepsilon] - \frac{|\bar{a}_-|}{\bar{k}}\} . \quad (33)$$

In expressions (32), (33) $\xi^- = \inf_{0\le u<\infty}\xi(u)$ is the extreme value of the process. The distributions of all other boundary functionals (in particular, those associated with the intersection of a fixed level) are expressed through the distribution of extremal functionals; $\gamma^-(x) = \xi(\tau^-(x)) - x$ is first overjump over $x$<0.

For of the overjump value of the negative level $\gamma^-$ we obtain an exponential distribution with parameter $c_1$ and moments $\left\langle (\gamma^-)^n \right\rangle = n!/c_1{}^n$. Also, the exponential distribution is valid for the magnitude of the positive level overjump. If $m$>0, then $\rho_-\big|_{s\to 0} \to \rho_- = bp_- > 0$, and the description is more cumbersome.

## 4. Explicit expressions for moments

In this section, the relationships written in the previous sections are reduced to a form that allows calculation of dependencies for specific physical models. Such calculations were performed for the four-state model from Refs. [1], [4]. The solution of the Lundberg equation (11) is given. In this section the solution of the third order equation is given, and in Appendix A - the fourth order. Inequalities for upper and lower bounds of means and variances are also written. The discussion of the connections between the theory of random processes and statistical physics is continued.

### 4.1. General expressions for average values and correspondence between risk theory and thermodynamics of trajectories

From inequality Eq. (3) and expression (5) we obtain, taking into account the change of variable:

$$\left\langle A(s) \right\rangle = -\partial \ln Z_t(s)/\partial s \le -t\partial\tilde{\Lambda}(s)/\partial s = \left\langle A_*(s) \right\rangle , \quad (34)$$

$$\left\langle A_*(s) \right\rangle = -t\partial\tilde{\Lambda}(s)/\partial s = t[\sum\nolimits_{x\ne y}\pi_x w_{xy} a_{xy} e^{-sa_{xy}} - \frac{q\left\langle a^2 \right\rangle_\pi s(2+5qcs/\varepsilon)}{\varepsilon(1+5qcs/\varepsilon)^2}] , \quad (35)$$

where $\left\langle A_*(s) \right\rangle$ is the maximum possible upper value of the mean $\left\langle A(s) \right\rangle$, Eq. (56), $s$=-$u$, the variable $u$ from Eq. (4). For $s$=0 and $a_{xy}$=$1$, inequality Eq. (34) becomes equality

$\langle k\rangle_{\pi}=\sum_{x\neq y}\pi_x w_{xy}=\langle a\rangle_{\pi}=\sum_{x\neq y}\pi_x w_{xy}a_{xy}=m$. Positive and negative values of $m$ are possible. When $s\neq 0$ the value $\langle A_*(s)\rangle$ can also take on positive and negative values.

The maximum possible upper value of the variance of value $A$ is equal to:

$$\langle D_{A*}(s)\rangle = t\partial^2\tilde{\Lambda}(s)/\partial s^2 = t[\sum_{x\neq y}\pi_x w_{xy}a^2{}_{xy}e^{-sa_{xy}} + \frac{2q\langle a^2\rangle_{\pi}}{\varepsilon(1+5qcs/\varepsilon)^3}],\quad \langle D_A(s)\rangle \leq \langle D_{A*}(s)\rangle. \quad (36)$$

It can be shown that in the general case the inequalities for upper and lower bounds of means and variances are valid:

$$\bar{a}(s)\leq \bar{a}_*(s),\quad D_a(s)\leq D_{a*}(s),\quad \bar{\tau}_*(\gamma)\leq \bar{\tau}(\gamma),\quad D_{\tau}(\gamma)\leq D_{\tau*}(\gamma), \quad (37)$$

where $D_a(s)$ is the variance of the values $a$, $D_\tau(\gamma)$ is the variance of the $\tau$ values of *FPT*, $\bar{a}(s)=\boldsymbol{E}a(s)\simeq\langle A(s)\rangle/t$, $\bar{\tau}(\gamma)=\boldsymbol{E}\tau(\gamma)$. The values $\bar{\tau}_*(\gamma)$ are less $\bar{\tau}(\gamma)$ because it corresponds to the cumulant $\tilde{\Lambda}$ with maximum fluctuations when $\bar{\tau}(\gamma)$ are minimum. Let's show it. Since $g(\gamma)\leq g_*(\gamma)$, where $g(\gamma)=\ln\boldsymbol{E}(e^{-\gamma\tau})$, $g_*(\gamma)=\ln\boldsymbol{E}_*(e^{-\gamma\tau})$, $\ln\boldsymbol{E}_*(e^{-\gamma\tau})$ is the logarithm of the characteristic function, which, in accordance with Eq. (8), the inverse of the function $\tilde{\Lambda}$, then $-g_*(\gamma)\leq -g(\gamma)$, $\langle\tau(\gamma)\rangle\sim-\partial g(\gamma)/\partial\gamma$, $\langle\tau_*(\gamma)\rangle\leq\langle\tau(\gamma)\rangle$. If the first two inequalities in Eq. (37) can be obtained from the results of work [1], then the third and fourth inequalities in expression (37) represent an original result.

From Eq. (4), using relations (5), the average values of $A$ are determined. It's more difficult to get expressions for *FPT*. But to find *FPT*, you need to know the partition function Eq. (6) and the cumulant $g(\gamma)$ from Eq. (7). For generatrix (or *MGF*):

$$T(s,x)=\boldsymbol{E}[e^{-s\tau^+(x)},\tau^+(x)<\infty]=P\{\xi^+(\theta_s)>x\},\ \tau^+(x)=\inf\{t>0:\xi(t)>x\}$$

in works [57, 65] an integral equation with the solution close to Eq. (25) is derived:

$$T(s,x)=q_+(s)e^{-\rho_+(s)x},\quad q_+(s)=1-p_+(s),$$

where the values of $\rho_+(s)$, $p_+(s)=\rho_+(s)/c_1$ defined above.

For expression (25) for $x>0$ by $g(\gamma)=\ln\boldsymbol{E}[e^{-\gamma\tau^+(x)},\tau^+(x)<\infty]/x,\quad x=K$, we obtain:

$$g(\gamma)=-\rho_+(\gamma)+\ln[1-\rho_+(\gamma)/c_1]/x. \quad (38)$$

Determining *FPT* for a cumulant of the form Eq. (4) turns out to be more difficult than for the cumulant of the form Eqs. (19), (20). It can be seen from expression (38) that the second relation in Eq. (24) is not satisfied, and the term $\ln[1-\rho_+(\gamma)/c_1]/x$ is added to $-\rho_+(\gamma)$. Relations (24) are valid for the cumulant of the form Eqs. (19), (20). Also, expression (24) is satisfied for large values of $x=K$, and it is this situation of *LD* that is considered in the thermodynamics of trajectories.

To determine the cumulant $g(\gamma)$ for *FPT*, it is necessary to solve equation (11) from the form of $k(r)$ Eq. (4) with $u=r$ and find the roots $\rho_{\pm}$. As noted in work [38] and in Eqs. (5), (12), the partition function Eqs. (2)-(4) corresponds to the characteristic function of a random process, the cumulant $\theta(\gamma)$ from expression (5) corresponds to the cumulant $k(r=-\gamma)$ from expression (11), and the root $r_s>0$ Lundberg equation (11), which corresponds to the thermodynamics of trajectories relation (8), corresponds to the cumulant $-g(\gamma)$. However, in our case, equation (11) for cumulant Eq. (4) represents a complex transcendental equation.

Let us note the differences between the approaches of the theory of random processes and the thermodynamics of trajectories. So, for the $x$-ensemble, the fixed value of the value of $K$, which is reached in a random time *FPT*, is positive in the thermodynamics of trajectories, and in the theory of random processes the value of the limit *x* achieved by a random process can be both positive and negative. Accounting $1/c_1$ in Eq. (39) leads to the replacement in expression (41) for the average *FPT* of *x* by the factor $[1/c_1 + x(1-\rho_+/c_1)]$ in $\left\langle \tau^+(x) \right\rangle_{\gamma=0}$ Ref. [38]. If $1/c_1 = 0$ this multiplier is equal to *x*, as in Eq. (41).

The thermodynamics of trajectories does not take into account the processes of reaching negative levels. For *m*<0, $\rho_+ = \rho_+(s=0) \neq 0$, but $g(\gamma=0)=0$. In the thermodynamics of trajectories, the case *m*>0 is considered, when $\rho_+(s=0)=0$, and the case *m*<0 is not considered. The results obtained in Sections 2 and 3 for *m*<0 is not written using the thermodynamics of trajectories. The theory of random processes is a more complete theory that describes more possible physical situations than the thermodynamics of trajectories.

### 4.2. Characteristic functions and averages for *x*>0

For values *x*>0 (corresponding to $K>0$), we rewrite expression (25) by *u*=0 in the form:

$$\boldsymbol{E}[e^{-\gamma\tau^+(x)}, \tau^+(x)<\infty] = (1-\rho_+(\gamma)/c_1)e^{-\rho_+(\gamma)x}, \ x>0, \tag{39}$$

where the parameter *s* is replaced by *γ*.

Comparing, as above, the moment generating function Eq. (7), the Laplace transform of the distribution *FPT* with the characteristic function *FPT* of the form Eq. (39), we determine the average *FPT* values of level achievement in the form corresponding to the relationships used in statistical physics for the logarithm of the partition function, as in Eq. (15):

$$\left\langle \tau^+(x) \right\rangle_\gamma = -\frac{\partial \ln \boldsymbol{E}[e^{-\gamma\tau^+(x)}, \tau^+(x)<\infty]}{\partial\gamma} = \frac{\partial\rho_+(\gamma)}{\partial\gamma}[x + \frac{1}{c_1(1-\rho_+(\gamma)/c_1)}], \ x>0. \tag{40}$$

To determine the explicit dependence on the parameter *γ* in Eq. (40), it is necessary to solve equation (11) with function Eq. (4) (or other functions, for example Eqs. (A4), (45)) and determine the explicit form of the function $\rho_+(\gamma)$. Expression (27) was used above and the zero, first and second moments $\rho_+(\gamma)$ were determined. To determine the explicit form of the function $\rho_+(\gamma)$, you can use the Cramer-Lundberg, Renyi, De Wilder, diffusion and exponential approximations, Ref. [57]. A different approximation is used below.

Assuming in Eqs. (39), (40) $1/c_1 = 0$ we obtain:

$$\left\langle \tau^+(x) \right\rangle_\gamma = x\frac{\partial\rho_+(\gamma)}{\partial\gamma}, \tag{41}$$

which corresponds to Eq. (15) for *x*>0.

Transformation Eq. (8) corresponds to the Lundberg equation and the simple generatrix Eq. (17). In many cases, it is not possible to find an explicit form of the solution to equation (11). In such situations, asymptotic approximations such as the Cramer-Lundberg approximation [57] can be used. It seems promising to use expressions similar to Eq. (40). Typically, the average *FPT* value is sought by differentiating the characteristic function for *FPT*, after which the differentiation argument is set to zero. It is proposed to differentiate the logarithm of the characteristic function for *FPT*, as is done for the logarithm of the partition function Eq. (7).

A correspondence is established between the characteristic function for *FPT* Eq. (39) and partition function Eq. (7). Moreover, the parameter *γ* is not equal to zero after differentiation. Parameter *γ* plays the role of a physical field conjugate to a random thermodynamic parameter *FPT*, Refs. [12, 34-35]. This parameter in Refs. [34-35, 37-38] is associated with changes in the entropy in the system, in Refs. [35], [68] it is associated with flows in the system and entropy production. Therefore, the use of the results of the theory of random processes with the involvement of physical relationships, similar to the use of the partition function, the Legendre transformation, etc., seems promising.

In Refs. [34-35], the Laplace transform of the probability density $f(T_\gamma)$ of the *FPT* $T_\gamma$ distribution is the non-equilibrium part of the partition function $Z_\gamma = \int_0^\infty e^{-\gamma x} f(T_\gamma = x)dx$. In the thermodynamics of trajectories, Refs. [42, 48, 50], from Eq. (5) we obtain $\theta(s) = t^{-1} \ln Z_t(s)$, where the function $\theta(s)$ is considered as (negative) dynamic free energy per unit time, $Z_\tau(s) = \sum_K e^{-sK} P_\tau(K)$, Eq. (5), $t=\tau$, where $P_\tau(K)$ is the distribution of the all trajectories $\boldsymbol{X}_\tau$ of total time $\tau$ of dynamic activity $K$, defined as the total number of configuration changes per trajectory, Refs. [42, 48, 50]. This coincides for the case of positive processes $\xi(t) \geq 0$ with the definition of the process cumulant $\Psi(\alpha) = t^{-1} \ln \boldsymbol{E}(e^{i\alpha\xi(t)})$ Eq. (9) in the theory of random processes. For processes that, as in case Eq. (1), can also take negative values, the relationship between the characteristic function and the Laplace transform becomes more complicated.

The expressions for $\left\langle \tau^+(x) \right\rangle$ Eqs. (15) and (41) at $K=x$ coincides at $\gamma=0$. But, if we do not assume $\gamma=0$, but use relations of the form $\ln Z_K(\gamma) = Kg(\gamma)$, $\left\langle \tau^+(x) \right\rangle = -\partial \ln Z_K(\gamma) / \partial\gamma = -K\partial g(\gamma) / \partial\gamma$, and consider the parameter $\gamma$, as in works [33-34], as a field conjugate to *FPT*, then for relations (6)-(7) correspondence of the form is established:

$$\ln Z_K(\gamma) \to \ln \boldsymbol{E}[e^{-\gamma\tau^+(x)}, \tau^+(x) < \infty], \quad K \to x, \quad x > 0 .$$

If we put $\gamma=0$ in Eq. (40), then we get that the expression for the average *FPT* at $x>0$ is related to the expressions from Eqs. (39)-(40) as follows: $\left\langle \tau^+(x) \right\rangle_{\gamma=0} = \left\langle \tau^+(x) \right\rangle / E\{\tau^+(x) < \infty\}$. In Refs. [12, 33-34], this value $\left\langle \tau^+(x) \right\rangle_{\gamma=0}$ is interpreted as unperturbed average value of *FPT*. In real physical systems, there are always flows, entropy production and other effects, which are expressed through the parameter $\gamma$ and are reflected in expressions (39)-(40). In the general case, the system is in a non-equilibrium state; the parameter $\gamma=0$ is only in equilibrium. If in Eq. (40) compared to Eq. (41) the factor before $\partial\rho_+(\gamma) / \partial\gamma\big|_{\gamma=0}$ is different, $x$ is replaced by $x+1/c_1(1-\rho_+/c_1)$ (or $1/c_1+x$ for $m>0$), then in expressions (8), (40) in this factor a functional dependence of $\rho_+(\gamma)$ on $\gamma$ appears. However, expression (40) turns into Eq. (41) for *LD*, large values of $x$.

In the work [68], the proportionality of the parameter $\gamma$ to the entropy production $\sigma$ is shown from two sources. In one case, the distribution for *FPT*, which is included in the statistical distribution containing *FPT* as a thermodynamic parameter Refs. [34-35], is expanded in a series. In this case, the relation $\gamma = (\frac{2\Delta s}{D_{T_0}})^{1/2}$ is written, where $\Delta s$ is the change in entropy over time $T_0$ – the unperturbed mean of *FPT*, $D_{T_0}$ is the variance of the unperturbed *FPT*. If use *TUR* [3-8] according to which $2T_0^2/D_{T0} \sim T_0\bar{\sigma}$, where $\bar{\sigma}$ is mean the entropy production, then substituting this expression into the one written above with $\Delta s=\sigma T_0$, we obtain $\gamma \sim \bar{\sigma}$. In the second approach, the method of the nonequilibrium statistical operator Refs. [63-64] is used.

### 4.3. Characteristic functions and averages for $x<0$

Relations similar to expressions (6)-(7) for $x<0$ are written using expression (31) rewritten as:

$$\boldsymbol{E}[e^{-\gamma\tau^-(x)}, \tau^-(x)<\infty] = (1-\rho_-(\gamma)/b)e^{\rho_-(\gamma)x},\ x<0\,, \tag{42}$$

$$\left\langle \tau^-(x)\right\rangle_\gamma = -\frac{\partial \ln \boldsymbol{E}[e^{-\gamma\tau^+(x)}, \tau^+(x)<\infty]}{\partial\gamma} = \frac{\partial\rho_-(\gamma)}{\partial\gamma}[-x+\frac{1}{b(1-\rho_-(\gamma)/b)}],\ x<0\,. \tag{43}$$

In Appendix A and in expressions (26), (30), the behavior of cumulant Eq. (4) is modeled by an exponential distribution. However, calculations show that the exponential distributions used in Eqs. (26), (30), (A.2), (A.4) do not describe the approximations to the cumulant Eq. (4) as accurately as some other approximations.

### 4.4. Approximations used in calculations

The exponential distribution in Eqs. (26), (30), (A2), (A4) was chosen in order to analytically solve equation (11), which was done by solving the fourth-degree equation in Eq. (A.12). For other distributions chosen in the form of models of the behavior of process jumps Eq. (1), basically, transcendental equations for equation (11) are written. But the behavior of the average value of $A$ Eq. (35) in the case of choosing the exponential distribution in Eqs. (A.2)-(A.4) differs from the exact solution in the four-state model from works [1], [4]. Fig. 2 shows a comparison of the average values of $A$ for the exponential approximation $\left\langle a\right\rangle_{\exp}(s)$, the quadratic approximation $\left\langle a\right\rangle_{qu}(s)$ Eq. (45) in the expansion Eq. (44) of the cumulant Eq. (4) and the exact calculation $\left\langle a\right\rangle_{ex}(s)$ Eq. (35) with the cumulant Eq. (4). In addition to the model of exponential distribution of process jumps Eq. (1), other distributions were also considered: uniform distribution, $\chi^2$-distribution, and several others. The behavior of the uniform distribution (not shown in Fig. 2) is closest to the exact solution over large ranges of variation of the parameter $\gamma$. But in equation (11), the cumulant of this distribution also leads to the transcendental equation. And we are interested in the behavior of the cumulant in the interval $0\le u<\varepsilon/5qc$ ($0<u<0.37$ for the four-state model), considered in work [1] (Subsection 2.1). Expanding the exponential in the characteristic function of the uniform distribution into a series allows us to obtain a simple quadratic equation for expression (11), but leads to the behavior of the average value, which differs from the behavior of the exact solution. To obtain the final solution in analytical form, we start with the exact solution, expanding the exponential in expression (4) (at $s=-u$) into the series:

$$\theta(s=-u) = \sum\nolimits_{x\ne y}\pi_x\omega_{xy}(1-sa_{xy}+s^2a_{xy}{}^2/2-s^3a_{xy}{}^3/6+\dots)-\bar{k}+\frac{(q/\varepsilon)\left\langle a^2\right\rangle_\pi s^2}{1+(5qc/\varepsilon)s} =$$
$$= -s\left\langle a\right\rangle_\pi + s^2\left\langle a^2\right\rangle_\pi/2 - s^3\left\langle a^3\right\rangle_\pi/6+\dots+\frac{(q/\varepsilon)\left\langle a^2\right\rangle_\pi s^2}{1+(5qc/\varepsilon)s}\,. \tag{44}$$

Restricting expansion (44) to the quadratic term, we obtain an expression for the truncated cumulant of the form:

$$\theta_2(s=-u) = \sum\nolimits_{x\ne y}\pi_x w_{xy}(1-sa_{xy}+s^2a_{xy}{}^2/2)-\bar{k}+\frac{(q/\varepsilon)\left\langle a^2\right\rangle_\pi s^2}{1+(5qc/\varepsilon)s} = -s\left\langle a\right\rangle_\pi + s^2\left\langle a^2\right\rangle_\pi/2+\frac{(q/\varepsilon)\left\langle a^2\right\rangle_\pi s^2}{1+(5qc/\varepsilon)s}\,. \tag{45}$$

In Fig. 1 shows a comparison of the exact solution Eq. (4) and the quadratic approximation Eq. (45). Qualitatively, the behavior of the dependences coincides, differing in quantitative values (at large intervals of change in the parameter $u$), but coinciding in the interval $0<u<0.37$, which is set to work [1]. The advantage of the quadratic approximation compared to the exponential is the absence of the restrictions that exist in the exponential approximation.

Expressions for the average values of fluctuations Eq. (35) and their variances Eq. (36) are equal in the original expressions (4), (35), (36).

In Eqs. (35), (36) the first and second moments of $A$ are indicated depending on $s=-u$. The static average $\langle a \rangle_\pi$ and the second moment $\langle a^2 \rangle_\pi$, with the help of which the relative errors were written in work [1], are included in the right sides of Eqs. (35), (36), (44)-(47).

For the quadratic approximation Eq. (45):

$$\langle A_{*qu}(s,t) \rangle \simeq t\langle a \rangle_{*qu}(s) = t[\langle a(t) \rangle - s\langle a^2(t) \rangle + s^2\langle a^3(t) \rangle / 2 - \frac{(q/\varepsilon)\langle a^2 \rangle_\pi s(2+(5qc/\varepsilon)s)}{(1+(5qc/\varepsilon)s)^2}], \quad (46)$$

$$D_{A*qu}(s) = t[\langle a^2(t) \rangle - s\langle a^3(t) \rangle + s^2\langle a^4(t) \rangle / 2 + \frac{2(q/\varepsilon)\langle a^2(t) \rangle_\pi}{(1+(5qc/\varepsilon)s)^3}]. \quad (47)$$

The possibility of description with the expansion of the exponential in (4), when the cumulant has the form Eq. (44) is used here. Estimates show (Fig. 1) that these expansions, up to the quadratic term $k2(r)=\theta_2(u=r)$ Eq. (45) for the four-state model, agree with the exact solution $k(r)=\theta(u=r)$ Eqs. (4), (44). Calculations in Figs. 1-6 are made for the four-state model from Refs. [1], [4]. A current is defined by the values of the six coefficients $a_{x>y}=-a_{x<y}$ which we take in the range $a_{xy} \in [-1,1]$ with coefficients $a_{12}=0.9$, $a_{13}=-0.9$, $a_{14}=-0.9$, $a_{23}=0.9$, $a_{24}=-0.9$, $a_{34}=0.9$ from Eq. (1); $w_{12}=3$, $w_{13}=10$, $w_{14}=9$, $w_{21}=10$, $w_{23}=1$, $w_{24}=2$, $w_{31}=6$, $w_{32}=4$, $w_{34}=1$, $w_{41}=7$, $w_{42}=9$ and $w_{43}=5$ [4]; in Eq. (A.4) $\pi_i = 1/4, i = 1,2,3,4, \quad \alpha_+ = 0.36, \quad \alpha_- = 0.64$.

### 4.5. Explicit form of calculated expressions

In Eq. (55), solutions to equation (11) are obtained for a quadratic approximation Eq. (45) in expansion Eq. (44). From expressions (39)-(40), (42)-(43), (53) we obtain expressions for the means and variance of *FPT* achieving positive and negative levels:

$$\langle \tau^+ \rangle = \frac{\partial \rho_+(\gamma)}{\partial \gamma}(x + \frac{1}{c_1 - \rho_+(\gamma)}), \quad \rho_+(\gamma) = x_2(\gamma), \quad c_1 = \frac{\varepsilon}{5qc}, \quad x > 0, \quad (48)$$

$$\frac{\partial \rho_+(\gamma)}{\partial \gamma} = -\frac{1}{\sqrt{Q}}\frac{\partial Q}{\partial \gamma}\cos(\Phi - \frac{2\pi}{3}) + 2\sqrt{Q}\sin(\Phi - \frac{2\pi}{3})\frac{\partial \Phi}{\partial \gamma}, \quad \frac{\partial \Phi}{\partial \gamma} = -\frac{1}{3}\frac{1}{\sqrt{Q^3 - R^2}}(\frac{\partial R}{\partial \gamma} - \frac{3}{2}\frac{\partial Q}{\partial \gamma}\frac{R}{Q}),$$

$$\langle \tau^- \rangle = \frac{\partial \rho_-(\gamma)}{\partial \gamma}(x - \frac{1}{b_2 + \rho_-(\gamma)}), \quad \rho_-(\gamma) = -x_1(\gamma), \quad b_2 = m_{3-} = \frac{\langle k \rangle_\pi}{\langle a_- \rangle_{\pi-}}, \quad x < 0, \quad (49)$$

$$D_{\tau+} = \frac{\partial^2 x_2}{\partial \gamma^2}(x + \frac{1}{(c_1 - x_2)}) + (\frac{\partial x_2}{\partial \gamma})^2 \frac{1}{(c_1 - x_2)^2}, \quad x>0, \quad (50)$$

$$D_{\tau-} = -\frac{\partial^2 x_1}{\partial \gamma^2}(x - \frac{1}{(b_2 + x_1)}) - (\frac{\partial x_1}{\partial \gamma})^2 \frac{1}{(b_2 + x_1)^2}, \quad x<0. \quad (51)$$

The quantities $x_1$, $x_{2,3}$, roots of equation (11), written below in the form Eq. (52), $\Phi$, $R$, $Q$ are defined in Eq. (53). For the four-state model their values are equal:

$Q(\gamma)=10^{-2}(2.45684\gamma+1,91)$, $\quad R(\gamma)=10^{-2}(452.256\gamma+7,98)$,

$$\frac{\partial x_1(\gamma)}{\partial\gamma}=-\frac{\partial\rho_-(\gamma)}{\partial\gamma}=-\frac{1}{\sqrt{Q}}\frac{\partial Q}{\partial\gamma}\cos(\Phi)+2\sqrt{Q}\sin(\Phi)\frac{\partial\Phi}{\partial\gamma},\quad \frac{\partial\Phi}{\partial\gamma}=-\frac{1}{3}\frac{1}{\sqrt{Q^3-R^2}}\left(\frac{\partial R}{\partial\gamma}-\frac{3}{2}\frac{\partial Q}{\partial\gamma}\frac{R}{Q}\right),$$

$$\frac{\partial x_2(\gamma)}{\partial\gamma}=-\frac{1}{\sqrt{Q}}\frac{\partial Q}{\partial\gamma}\cos\left(\Phi-\frac{2\pi}{3}\right)+2\sqrt{Q}\sin\left(\Phi-\frac{2\pi}{3}\right)\frac{\partial\Phi}{\partial\gamma},$$

$$\frac{\partial^2 x_1(\gamma)}{\partial\gamma^2}=-\frac{1}{2\sqrt{Q^3}}\left(\frac{\partial Q}{\partial\gamma}\right)^2\cos(\Phi)+\frac{2}{\sqrt{Q}}\sin(\Phi)\frac{\partial Q}{\partial\gamma}\frac{\partial\Phi}{\partial\gamma}+2\sqrt{Q}\cos(\Phi)\left(\frac{\partial\Phi}{\partial\gamma}\right)^2+2\sqrt{Q}\sin(\Phi)\frac{\partial^2\Phi}{\partial\gamma^2},$$

$$\frac{\partial^2\Phi}{\partial\gamma^2}=\frac{1}{6}\frac{1}{\sqrt{(Q^3-R^2)^2}}\left(\frac{\partial R}{\partial\gamma}-\frac{3}{2}\frac{\partial Q}{\partial\gamma}\frac{R}{Q}\right)\left(3Q^2\frac{\partial Q}{\partial\gamma}-2R\frac{\partial R}{\partial\gamma}\right)+\frac{1}{2\sqrt{Q^3-R^2}}\frac{\partial Q}{\partial\gamma}\frac{1}{Q}\left(\frac{\partial R}{\partial\gamma}-\frac{\partial Q}{\partial\gamma}\frac{R}{Q}\right),$$

$$\frac{\partial^2 x_2(\gamma)}{\partial\gamma^2}=-\frac{1}{2\sqrt{Q^3}}\left(\frac{\partial Q}{\partial\gamma}\right)^2\cos\left(\Phi-\frac{2\pi}{3}\right)+\frac{2}{\sqrt{Q}}\sin\left(\Phi-\frac{2\pi}{3}\right)\frac{\partial Q}{\partial\gamma}\frac{\partial\Phi}{\partial\gamma}+2\sqrt{Q}\cos\left(\Phi-\frac{2\pi}{3}\right)\left(\frac{\partial\Phi}{\partial\gamma}\right)^2+2\sqrt{Q}\sin\left(\Phi-\frac{2\pi}{3}\right)\frac{\partial^2\Phi}{\partial\gamma^2}.$$

The original point is the ability to obtain expressions to achieve negative values.

In Ref. [37], the relations for correlations between dynamical activity and *FPT* were obtained. For correlations between value *A* and *FPT* these relations have a form:

$$D_{A\tau\pm}=D_A\frac{\langle\tau^\pm\rangle}{\langle A(t)\rangle}.$$

Expressions for $D_A$, $\langle\tau^\pm\rangle$, $\langle A(t)\rangle$ are given in Eqs. (35), (36), (44)-(51).

### 4.6. Solution of equation (11) and calculation results

Figure 1 shows a comparison of the cumulant *k(r)* of the form $\tilde{\Lambda}(r)$ Eq. (4) and the quadratic approximation *k2(r)* Eq. (45), calculated for the four-state model from Refs. [1], [4]. On Fig. 1a shows the full range of changes in the argument $0\le u<\varepsilon/5qc$, as adopted in Eq. (4), in Fig. 1b – negative part of the cumulants; in the interval $0<r<0.1857$ they have small negative values. What is significant is their almost complete coincidence.

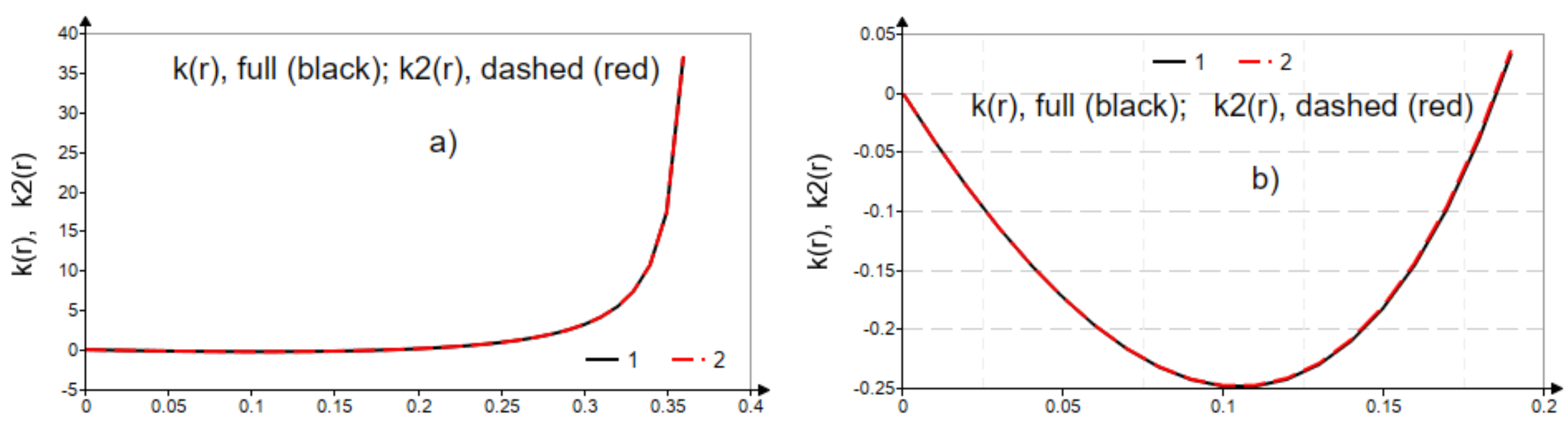

Fig.1. Cumulants *k(r)*, full, black, and *k2(r)*, dashed, red, in the ranges *r*: 0 - 0.37 (Fig. 1a), and 0 - 0.1857 (Fig. 1b), when $k(r)<0$, $0<r<0.1857$.

More complete agreement is given by expansions taking into account the terms $s^4$ and $s^6$. It can be seen that in the interval $0<u=r=-s<1/g_3$ the behavior of the expression for the cumulant *k2(r)* of the quadratic expansion of the form Eq. (45) is very close to the behavior of the exact solution Eq. (44) with the cumulant *k(r)*. At $0<r<0.1857$, the values of both cumulants are negative, Fig. 1b; at $0.1857<r<0.37$ – positive, Fig. 1a.

Comparison of expressions for *a(s)* ($\langle a\rangle_{qu}(s)$, $\langle a\rangle_{ex}(s)$, $\langle a\rangle_{\exp}(s)$) obtained from cumulant with exponentially modeled distributions of the form Eq. (A.2) ($\langle a\rangle_{\exp}(s)$) with the exact relation $\langle a\rangle_{ex}(s)$ from Eq. (4) and quadratic approximation $\langle a\rangle_{qu}(s)$ Eq. (46) is given in Fig. 2 for the interval *s*, $-1/g_3 < s < 0$, corresponding to the interval $0<u=-s < 1/g_3$, $g_3$ in Eq. (52) [1].

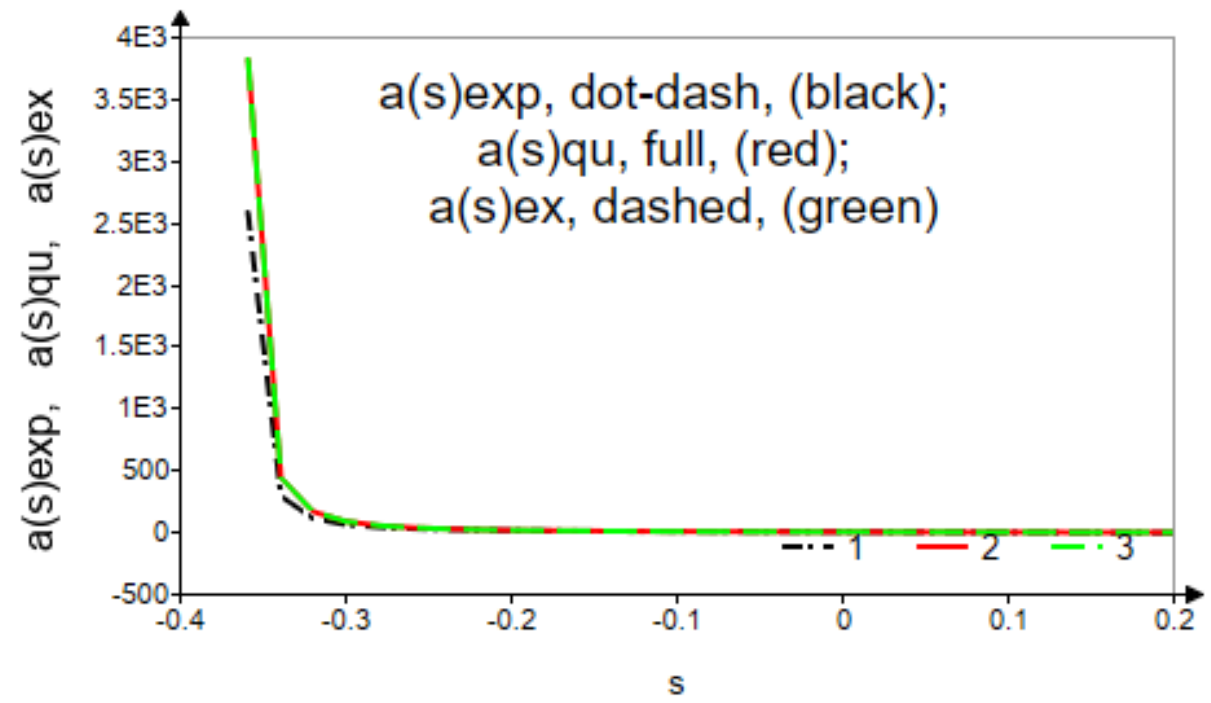

Fig. 2. Average values of size of fluctuations $\langle a(s,t)\rangle = \langle A(s,t)\rangle / t$ for exponential distribution $\langle a\rangle_{\exp}(s)$ (dot-dash, black), $\langle a\rangle_{qu}(s)$ (full, red), Eq. (46), quadratic approximation, and exactly solution of Eq. (4) $\langle a\rangle_{ex}(s)$ (dashed, green), Eq. (35).

From the calculation results in Fig. 2 it follows that the quadratic approximation corresponds better to the exact solution than the exponential model. Therefore, in further calculations, the quadratic approximation is used. In intervals with large values of *s*, the exact expression for $\langle a(s,t)\rangle$ corresponds to a uniform distribution for *f* from Eq. (A.2) and expressions from expansion Eq. (44) with even terms for *s* ($s^2$, $s^4$, $s^6$). Obtained using exponential and other distributions, for example, expansion from Eq. (44) with odd powers of *s*, the values $\langle a(s,t)\rangle = \langle A(s,t)\rangle / t$ over large intervals of variation in *s* differ significantly from the exact expression for $\langle a(s,t)\rangle$.

Let us consider equation (11) for $s=\gamma$ with cumulant *k2(r)* chosen in the form of a quadratic approximation Eq. (45) in expansion Eq. (44). Then equation (11) takes the form:

$$g_3\langle a^2\rangle_\pi r^3 + [\langle a\rangle_\pi g_3 - \langle a^2\rangle_\pi (1/2+q/\varepsilon)]r^2 - (\langle a\rangle_\pi + \gamma g_3)r + \gamma = 0, \quad g_3 = 5qc/\varepsilon = 1/c_1. \quad (52)$$

Solving cubic equation (52) using Vieta's trigonometric formula, work [66], we obtain three roots of this equation in the form:

$$x_1 = -2\sqrt{Q}\cos(\Phi) - [\langle a\rangle_\pi g_3 - \langle a^2\rangle_\pi (1/2+q/\varepsilon)]/g_3\langle a^2\rangle_\pi, \quad \Phi = \frac{1}{3}\arccos(R/\sqrt{Q^3}), \quad (53)$$

$$x_{2,3} = -2\sqrt{Q}\cos(\Phi \mp 2\pi/3) - [\langle a\rangle_\pi g_3 - \langle a^2\rangle_\pi (1/2+q/\varepsilon)]/g_3\langle a^2\rangle_\pi,$$

$$Q = \{[\langle a\rangle_\pi g_3 - \langle a^2\rangle_\pi (1/2+q/\varepsilon)]^2 + 3g_3\langle a^2\rangle_\pi (\langle a\rangle_\pi + \gamma g_3)\}/9(g_3\langle a^2\rangle_\pi)^2,$$

$$R = \{2[\langle a\rangle_\pi g_3 - \langle a^2\rangle_\pi (1/2+q/\varepsilon)]^3 +$$

$$+9g_3\langle a^2\rangle_\pi [\langle a\rangle_\pi g_3 - \langle a^2\rangle_\pi (1/2+q/\varepsilon)](\langle a\rangle_\pi + \gamma g_3) + 27(g_3\langle a^2\rangle_\pi)^2\gamma\}/54(g_3\langle a^2\rangle_\pi)^3.$$

The calculation using these expressions, shown in Fig. 3 for the four-state model, leads to the fact that only the root $x_2$ satisfies the necessary conditions for $x_2=\rho_+$: the root $x_2$ is in the interval $0<x_2=r=u=-s<1/g_3$. The root $x_3$ lies above the interval $0<u=-s<1/g_3$, which is considered in work [1], and root $x_1$=-$\rho_-$ is negative. The $x_1$ root is used below to determine the *FPT* of negative values: $x_2(\gamma)=\rho_+(\gamma),\ x_1(\gamma)=-\rho_-(\gamma)$. Thus, in accordance with the general theory, the positive and negative roots of equation (11) were obtained, satisfying the requirements of the model under study.

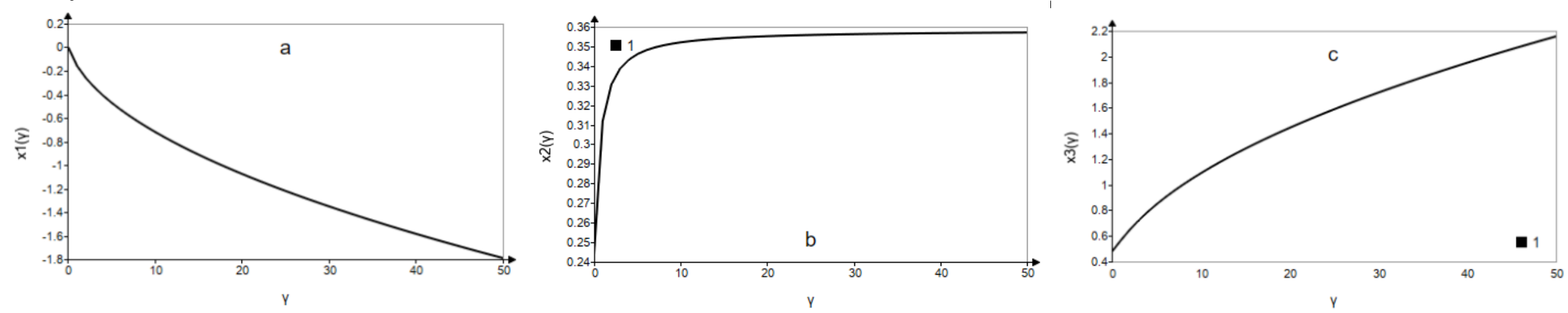

Fig. 3. Behavior of the three roots Eq. (53) of equation (52) depending on the parameter $\gamma$; $x_1$ (*a*), $x_2$ (b), $x_3$ (c) in the range of $\gamma$: 0-50.

Fig. 3 shows the behavior of $x_j$, $j$=1,2,3 values, depending on the parameter $\gamma$, and Fig. 4 shows the behavior of the average *FPT* $\tau(+)(\gamma)=\langle\tau^+(x,\gamma)\rangle$ (40), (48) for $x_2=\rho_+$ and the behavior of the *FPT* dispersion in the same range of $\gamma$.

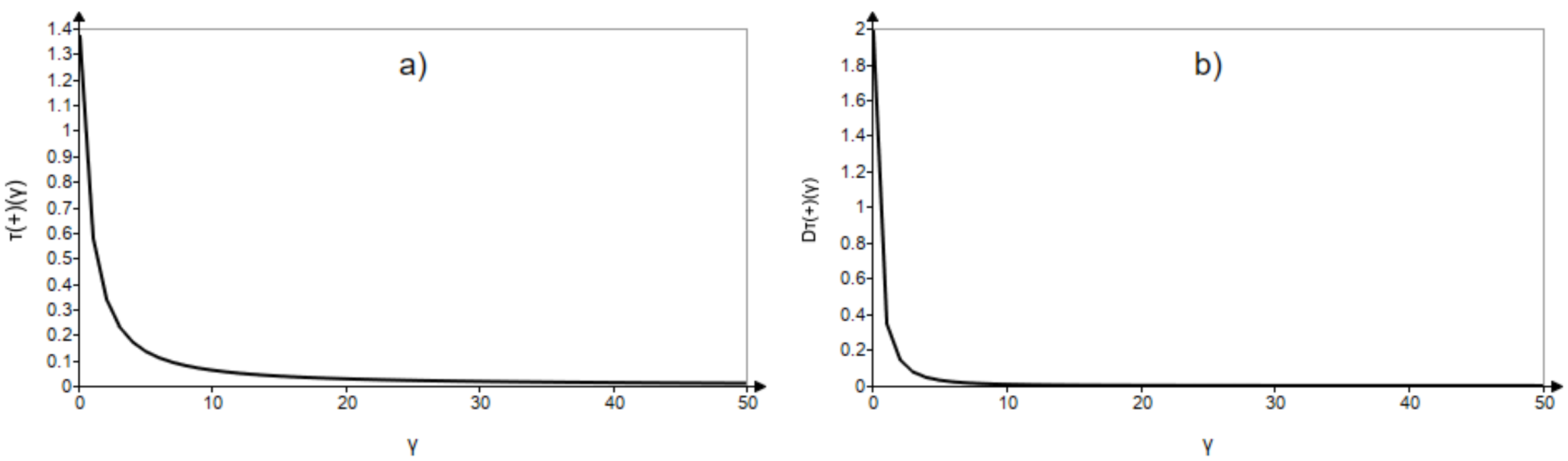

Fig. 4. Behavior of the average *FPT* $\tau(+)(\gamma)=\langle\tau^+(x,\gamma)\rangle$ (Fig. 4a)) Eqs. (40), (48) at $x_2=\rho_+(s)$ for the size of fluctuations reaching a positive level $x$=1 depending on the parameter $\gamma$ in the range of $\gamma$: 0-50. Dispersion of *FPT* (Fig. 4b)) $D_\tau(+)(\gamma)=D_T(\gamma)$ Eq. (50) reaching a positive level $x$=1 for the size of fluctuations.

Let us evaluate numerically the conclusions of Section 4.2. In Section 4.2 it was noted that in expression (40) in the factor $x+(c_1-\rho_+)^{-1}$ for large $x$, the second term can be neglected. In work [1] shows that $\varepsilon/5q\le 1.5$. For the four-state model $c$=0.9. Then $c_1=\varepsilon/5qc\le 1.67$, $\rho_+\simeq 0.35$, $(c_1-\rho_+)^{-1}\approx 0.76$ (in calculations it was assumed $c_1$=0.37). In the calculations, $x$=1 was taken, which is comparable to 0.76. But, if we take $x$=10 (or 100), then 0.76<<$x$. One more example. On the right side of expression (38) there are two terms: $-\rho_+(\gamma)+\ln[1-\rho_+(\gamma)/c_1]/x$. Numerical estimates give: -0.35-0.23/$x$. For $x$>10, the second term can be neglected, and expression (38) becomes Eq. (24). The general conclusions are confirmed by calculations.

The behavior of the average *FPT* value reaching a positive level $x$=1 for the size of fluctuations at the root $x_2=\rho_+$(s) Eq. (53) is written in expression (48) and is shown in Fig. 4a). Note that level $x$ can be arbitrary. The behavior of the *FPT* dispersion reaching a positive level Eq. (50) at $x$=1 is shown in Fig. 4b).

From relation (3) $Z_{v,t}(u)\le C(v)e^{t\tilde{\Lambda}(u)}$ we obtain:

$$\log Z_{v,t}(u) \le \log C(v) + t\tilde{\Lambda}(u). \tag{54}$$

For non-negative $u \in R$ valid expansion, work [1]:

$$\log Z_{\pi,t}(u) = \langle a \rangle_\pi tu + \sigma_\pi^{\ 2}(t)u^2/2 + o(u^2) \le \langle a \rangle_\pi tu + t\langle a^2 \rangle_\pi u^2(1+2q/\varepsilon)/2 + o(u^2) \simeq t\tilde{\Lambda}(u), \tag{55}$$

where $\sigma_\pi^{\ 2}(t)$ is the variance of the value *A(t)*, if $\pi$ is the initial distribution [1]. For derivatives of inequality Eq. (54) from Eq. (2) we obtain:

$$Z_{\pi,t}(u) \simeq \boldsymbol{E}[e^{uA(t)}], \quad \frac{\partial \ln Z_{\pi,t}(u)}{\partial u} \simeq \frac{\boldsymbol{E}[A(t)e^{uA(t)}]}{\boldsymbol{E}[e^{uA(t)}]} = \langle A(t) \rangle_u = \langle a \rangle_{\pi,u} t \le t\frac{\partial\tilde{\Lambda}(u)}{\partial u}, \quad \langle a \rangle_{\pi,u} \le \langle a \rangle_{*\pi,u},$$

$$\frac{\boldsymbol{E}[A(t)^2 e^{uA(t)}]}{\boldsymbol{E}[e^{uA(t)}]} - (\frac{\boldsymbol{E}[A(t)e^{uA(t)}]}{\boldsymbol{E}[e^{uA(t)}]})^2 = D_A(t,u), \quad \sigma^2_{\ v,u}(t) \le t\frac{\partial^2\tilde{\Lambda}(u)}{\partial u^2} = \sigma^2_{\ \pi,u}(t). \tag{56}$$

In [1] it is shown that $\sigma^2_{\ \pi}(t) \le t\langle a^2 \rangle_\pi (1+\frac{2q}{\varepsilon})$.

The exact expression for the variance $\sigma^2_{\ *\pi,u}(t) = t\frac{\partial^2\tilde{\Lambda}(u)}{\partial u^2}$ of *A(t)* for the four-state model is:

$$\sigma^2_{\ *\pi,u}(t) = 0.81[6e^{0.9\cdot u} + 10.75e^{-0.9\cdot u}] + 16.281/(1-2.74\cdot u)^3. \tag{57}$$

For expansion Eq. (44) in the quadratic approximation Eq. (45), the same expression is equal to:

$$\sigma^2_{\ *qu,\pi,u}(t) = 13.5675 - 3.46u + 5.49u^2 + 16.281/(1-2.74u)^3. \tag{58}$$

The dispersion $\sigma^2_{\ *qu,\pi,u=-s}(t)$ =$D_a(s)$ at $u$=-$s$ of the value *A(t)* in the quadratic approximation Eq. (58) and in the exact case $D_aex(s)$=$\sigma^2_{\ *\pi,u}(t)_{u=-s}$ Eq. (57) is shown in Fig. 5.

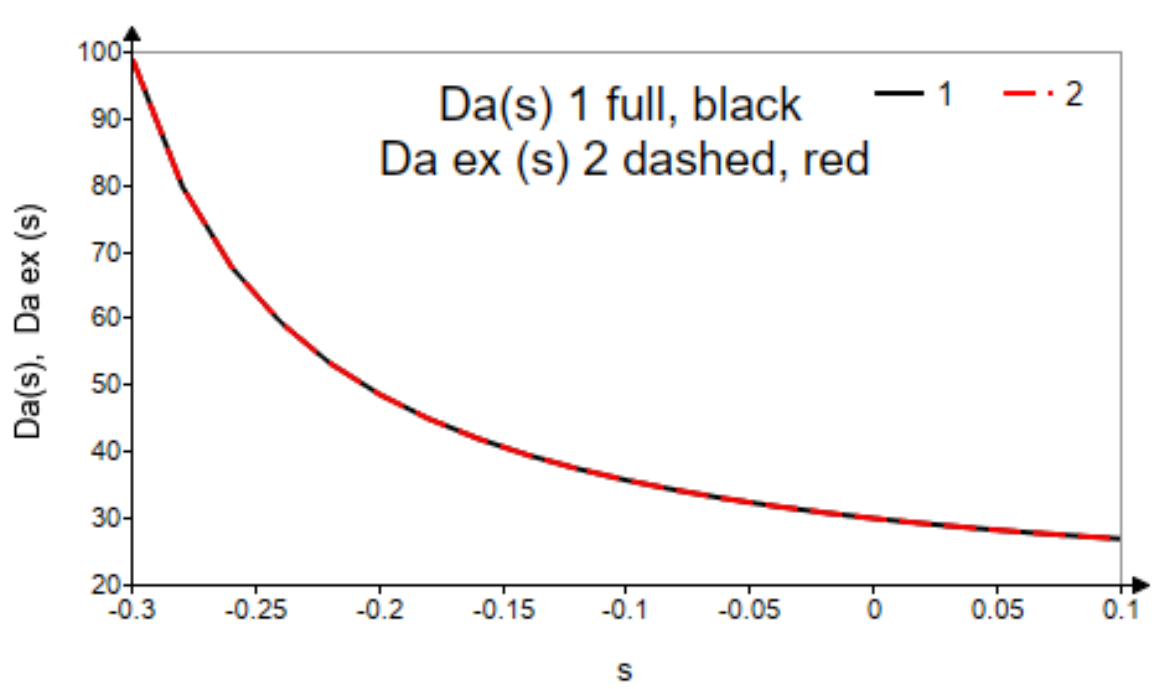

Fig.5. Dispersion of the fluctuation size in the quadratic approximation $D_a(s)$=$\sigma^2_{\ *qu,\pi,s=-u}(t)$ (full, black) Eq. (58) and in the exact case $D_aex(s)$=$\sigma^2_{\ *\pi,u}(t)_{u=-s}$ Eq. (57) (dashed, red).

The variance of the quadratic approximation of the fluctuation size equation (58) coincides with the variance of the exact solution of equation (57) for $s$<0, $u$>0. The same applies to taking into account the cubic term in expansion Eq. (44), exponential model Eq. (A.4), the behavior of which is very close to the quadratic approximation.

For $x$<0, $\rho_-(\gamma) = -x_1(\gamma)$, the average $\langle \tau^-(x,\gamma) \rangle$ and variance of *FPT* at $x$=-1 are equal to Eqs. (49), (51) and are shown in Fig. 6.

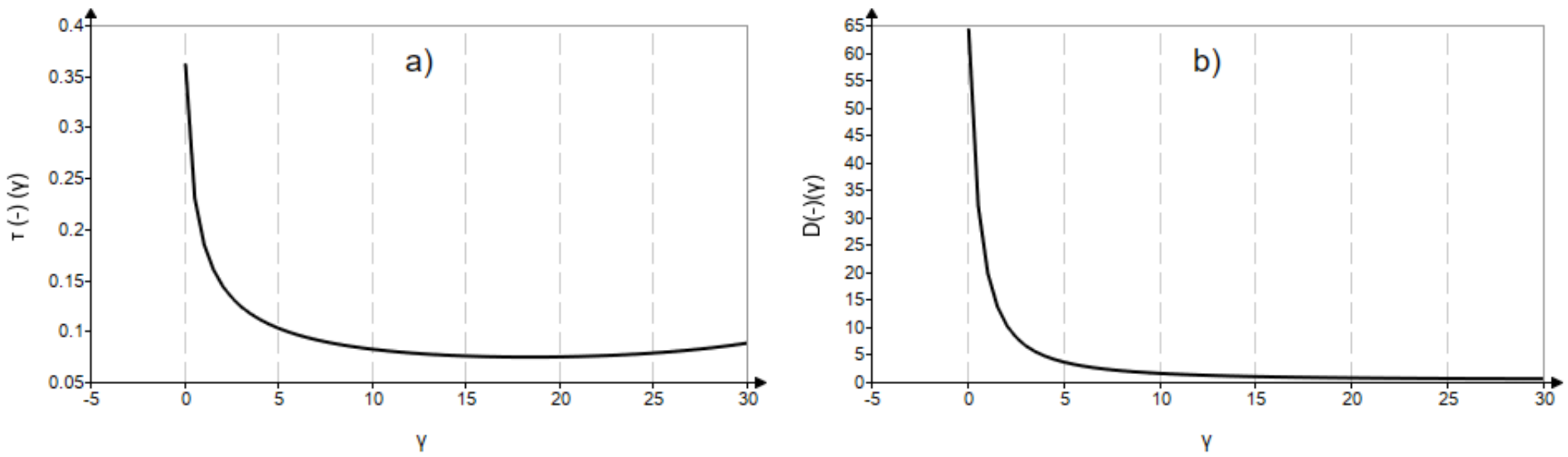

Рис. 6. Average *FPT* value $\tau(-)(\gamma)=\langle \tau^{-}(x,\gamma)\rangle$ (Fig.6a)) for $x<0$ at $x=-1$ Eq. (49); *FPT* variance $D(-)(\gamma)=\langle D^{-}(x,\gamma)\rangle$ (Fig.6b)) for $x<0$ at $x=-1$, Eq. (51).

The growth of the curve of the average *FPT* (Fig.6a)) value at $\gamma=30$ is caused by the approach of the singularity at $\gamma=44.3$

In expression (42), the condition $\rho_{-}(\gamma)/b_2<1,\ b_2=\langle k\rangle_{\pi}/\langle a_{-}\rangle_{\pi-}\approx 1.73$ for the four-state model must be satisfied. From Fig. 3a it is clear that to fulfill this condition, the values of the parameter $\gamma$ must be $\gamma\leq 45$, since for $\gamma>45$, $\rho_{-}(\gamma)=-x_1(\gamma)>1.73$.

## 5. Conclusion

The article obtained the average values and variances of general upper bounds on the size of fluctuations and of any linear combination of fluxes (including all time-integrated currents or dynamical activities) for continuous-time Markov chains. But the random variables *A* themselves can reach arbitrary values *x* (with certain probabilities), exceeding the average values. Therefore, in the expressions for the average values and dispersion of *FPT* on the size of fluctuations *A*, arbitrary values of the levels of achievement *x* appear, both positive and negative. It is possible, as in work [37], to express the average values and variances of general upper bounds on the size of fluctuations *A* and general upper bounds of *FPT* through changes in entropy $\Delta s$.

This paper follows papers [1] and [2]. However, paper [1], which introduces i*TUR*, does not investigate *FPT*. And in paper [2], the authors are interested in a stochastic process that counts the number of jumps that occurs in a stochastic trajectory up to time *t*. That is, an integer-valued and non-negative process. This paper examines the possibilities of achieving negative levels for the process *A(t)*, which characterizes the size of the fluctuations limited from above.

The moments of the fluctuation size *A* and their *FPT* provide complete and direct information about the size and behavior of the fluctuations, as does the rate limiting function associated with the dynamical entropy or the relative error $\epsilon_A$ for *A*, work [1]. Finding *FPT* for process Eq. (1) with cumulant Eq. (4) is a nontrivial task, Ref. [38]. Perhaps more interesting than the moments of fluctuation size are the *FPT* achievements of positive and negative levels. Just as *FPT*s have a large number of applications in a wide variety of fields, Refs. [10–12], upper bounds for *FPT*s should prove important in a wide variety of applications. From Figs. 4, 6 it is clear that $\langle \tau^{\pm}\rangle$ quickly decrease with increasing $\gamma$. In the papers [34-35] a statistical distribution is introduced and considered that contains the internal energy of the system and the *FPT* as random thermodynamic variables. The parameter $\gamma$ is present in this distribution as a thermodynamic parameter conjugate to the *FPT*. At $\gamma=0$ the system is in equilibrium with the Gibbs distribution. As the parameter $\gamma$ increases the system moves away from equilibrium. The parameter $\gamma$ can be considered as a measure of distance from equilibrium. As one moves away from equilibrium, fluctuations increase and their *FPT* decreases.

The dependences $\langle \tau^{\pm} \rangle$ on $x$ have the form $\langle \tau^{\pm} \rangle = Ax + B$. You can consider in detail the dependencies $\langle \tau^{\pm} \rangle$ on $\langle a \rangle_{\pi}$ and other parameters. The parameter $\gamma$ reflects external influences on the system (along with internal ones). From Figs. 4, 6 it is clear that $\langle \tau^{\pm} \rangle$ do not increase when exposed to $\gamma$. An increase $\langle \tau^{\pm} \rangle$ for process Eq. (1) with cumulant Eq. (4) is possible with an increase in the level $x$, as well as with a corresponding change in such internal process parameters as $\langle a \rangle_{\pi}$, $\langle a^2 \rangle_{\pi}$, $\bar{k}$, $c$, $q/\varepsilon$. In Ref. [1], it is noted that the inverse *TUR* captures the increase of fluctuations close to a dynamical phase transition via its dependence on $q/\varepsilon$. It is also stated in work [1] that the spectral gap of a (symmetrized) generator is used as input for the boundaries. Further approximations may also allow inverse *TUR*s to be formulated in terms of operationally available quantities. The results obtained can be linked to other articles. For example, to moment generating function Eq. (2) and cumulant function Eq. (4) you can apply the fluctuation relation obtained in Ref. [67] and write a constraint on the average of observables.

It is worth noting the relationship and interpenetration of methods of the theory of random processes and statistical physics carried out in the article. The article notes important analogies, as well as differences between the methods and approaches thermodynamics of trajectories and the theory of random processes, in particular, the theory of risk. This article concludes that the theory of random processes is a more complete theory that describes more possible physical situations than the thermodynamics of trajectories. Thus, the characteristic function *FPT* of the form Eqs. (38)-(39), (42) is used, obtained using the theory of random processes, works [38], [57]. But this characteristic function is compared with a partition function of the form Eq. (7) and in Eqs. (40)-(41), (43), (48)-(49) differentiation is carried out not of the characteristic function, but of its logarithm, as in statistical physics. Moreover, after differentiation, the argument of the characteristic function is not set equal to zero, as is done in the theory of random processes, where this argument plays an auxiliary role. In expressions (39), (42), the argument of the characteristic function is compared with the physical field, which is equal to zero only in the equilibrium case. There is a mutual influence on each other of the methods and approaches of statistical physics, in particular, thermodynamics of trajectories and the theory of random processes. An important point made in this article is the possibility of considering *FPT* reaching negative levels.

A significant place in the article is devoted to the connections and mutual influence of the theory of random processes and statistical physics, risk theory and thermodynamics of trajectories. Let us write down several general conclusions:

1). Probabilities $\boldsymbol{E}(\tau(x) < \infty)$ are written explicitly (for example, Eq. (29)), which, as a rule, are not taken into account in statistical physics.

2). In the thermodynamics of trajectories $g(\gamma = 0) = 0$, and for the case $m$<0, as shown in the article, the function $\rho_{+} = \rho_{+}(\gamma = 0) \neq 0$, where $\rho_{+}(\gamma)$ corresponds to the function $-g(\gamma)$; or ratio $-g(\gamma) = -\ln \boldsymbol{E}[e^{-\gamma\tau^{+}(x)}, \tau^{+}(x) < \infty] / x$. However, this depends on the sign of the mean value of the cumulant *k(r)* in Eq. (11), the value $m = \boldsymbol{E}\xi(1)$.

3). The characteristic function and moments of reaching negative levels $x$<0 is obtained.

4). Average values $\langle \tau \rangle$, which in statistical physics are expressed through a partition function $-\partial \ln Z_K(\gamma)/\partial\gamma$, are proposed to be expressed through a characteristic function $-\partial \ln \boldsymbol{E}[e^{-\gamma\tau^+(x)}, \tau^+(x)<\infty]/\partial\gamma$, $\gamma \neq 0$, although the definition $-\partial \boldsymbol{E}[e^{-\gamma\tau^+(x)}, \tau^+(x)<\infty]/\partial\gamma\big|_{\gamma=0}$ is used in the theory of random processes.

5). After differentiation by $\gamma$, when determining the average values, it is not necessary to assume $\gamma = 0$, since this corresponds to the equilibrium case, and in non-equilibrium situations $\gamma \neq 0$.

6). A combination of approaches from statistical physics and the theory of random processes is proposed, for example, the use of the Legendre transform for characteristic functions and cumulants of random processes.

7). A correspondence is established between the characteristic functions of the theory of random processes and the partition function $Z_K(\gamma)$ of statistical physics, between the cumulant $\theta(s)$ corresponding to the dynamic free energy unit time and the cumulant of the random process $\Psi(\alpha)$, Eq. (9).

8). The resulting correspondence between the level $x$ of achieving *FPT* and the amount of activity $K$ is possible only for $x$>0, since $K$>0.

9). The important role of risk theory in describing *FPT* processes with cumulant Eq. (4) is shown.

10). For cumulant Eq. (4), relation (24) $g(\gamma) = -\rho_+(\gamma)$ is supplemented, taking the form $g(\gamma) = -\rho_+(\gamma) + \ln[1-\rho_+(\gamma)/c_1]/x$, Eq. (38). However, expression (24) is satisfied for large values of $x$=$K$, and it is this situation of *LD* that is considered in the thermodynamics of trajectories. The same thing happens with *LD* with expressions (40) and (41).

11). Relations of the form (38)-(40) can be used to determine the *FPT* for achieving small levels of $x$, which are poorly defined in the situation of *LD* that is considered in the thermodynamics of trajectories.

Some of these conclusions are confirmed by calculations.

It was shown in Refs. [37]-[38], [71] that the argument $\gamma$ of the characteristic function, which in a statistical distribution containing the random energy $u$ and *FPT* $T_\gamma$ and conjugate $T_\gamma$, Refs. [34], [35], is equal to

$$\gamma = -\beta \partial\langle u\rangle/\partial\langle T_\gamma\rangle_{s_\gamma} = \beta(\partial\langle u\rangle/\partial t)_{s_\gamma}\,\chi_{tT_\gamma}, \text{ where } \chi_{tT_\gamma} = \begin{cases} 1, \ t \in \langle T_\gamma\rangle \\ 0, \ t \notin \langle T_\gamma\rangle \end{cases}, \ \partial t = -\partial\langle T_\gamma\rangle, \ \beta = \frac{1}{T} \text{ is the}$$

reverse temperature; $s_\gamma$ is the internal entropy; the quantity $\gamma$ is related to of the energy $u$ flux $q$ (or to the flux of another physical quantity, for example, the dynamic activity $K$ in Ref. [37]) and to the entropy production; $\gamma = -\beta \dfrac{\partial\langle u\rangle}{\partial\langle T_\gamma\rangle_{|s_\gamma}} = \beta \dfrac{\gamma - \langle u\rangle/D_{uT_\gamma}}{\beta - \langle u\rangle/D_u}$, where $D_u$ is dispersion of the energy $u$, $D_{uT\gamma}$ is a correlation between the parameters $u$ and $T_\gamma$. Just as the parameter $\beta$ is not equal to zero, so is the parameter $\gamma$ not equal to zero.

The results obtained in this article are important in many physical applications. For example, in Ref. [8] a thermodynamic uncertainty relation of the form of inequality $\langle\tau\rangle_A/\sqrt{\mathrm{var}(\tau)} \leq \sqrt{K_A}$ was obtained. In the notation of this article $\langle\tau\rangle_A = \tau_{\gamma=0}$, $\mathrm{var}(\tau) = D_\tau\big|_{\gamma=0}$, $K_A = K_0$. The results of this article give the opposite inequality for upper bound of the form: $\langle\tau\rangle_A/\sqrt{\mathrm{var}(\tau)} \geq \langle\tau\rangle_{*A}/\sqrt{\mathrm{var}_*(\tau)}$, where $\langle\tau\rangle_{*A} = \tau_{*\gamma=0}$ and $\mathrm{var}_*(\tau) = D_{*\tau}\big|_{\gamma=0}$ are the mean and variance of *FPT* obtained using cumulants $\tilde{\Lambda}$ (4).

## Appendix A. Modeling cumulants with exponential distribution.

As in Ref. [38], we rewrite Eq. (4) in the form Eq. (A.1). We proceed from the cumulant Eq. (4), written in the form:

$$k(r) = \frac{\lambda_1 r^2}{c_1 - r} + \sum_{x\neq y} \bar{k} f_1 (e^{ra_{xy}} - 1), \tag{A.1}$$

$\lambda_1 = \dfrac{\langle a^2\rangle_\pi}{5c}$, $c_1 = \dfrac{\varepsilon}{5cq}$, $\bar{k} = \sum_{x\neq y}\pi_x w_{xy} = \langle k\rangle_\pi$, $f_1 = \dfrac{\pi_x w_{xy}}{\bar{k}}$, $\dfrac{\partial k(r)}{\partial r}\Big|_{r=0} = m = \langle a\rangle_\pi = \sum_{x\neq y}\pi_x w_{xy} a_{xy}$,

$q = \max_x w_{xx}$, $c = \max_{x\neq y}|a_{xy}|$, $2q/\varepsilon \geq 1$.

Let us make an assumption about the form of the function $f_1$ from Eq. (A.1). The summation in Eq. (A.1) is replaced by integration over $a_{xy}$. Let us assume that $\pi_x w_{xy}$ depend on $a_{xy}$ and approximate this dependence of the density of the distribution function $f_1$ in Eq. (A.1) by an exponential distribution of the form:

$$f_{1+}(a_{xy}) = m_{3+}e^{-m_{3+}a_{xy}},\ a_{xy} > 0;\quad f_{1-}(a_{xy}) = m_{3-}e^{m_{3-}a_{xy}},\ a_{xy} < 0;\quad m_{3\pm} = \bar{k}/\left|\langle a\rangle_{\pi\pm}\right|. \tag{A.2}$$

The choice of exponential distribution for the distribution function $f_1$ in Eq. (A1) corresponds to the approximation often used in risk theory Ref. [57].

Values $a_{xy}$ take both positive and negative values. Let us divide the sum in Eq. (A.1) into two parts: one with positive values $a_{xy}$, the second with negative values:

$$\sum_{x\neq y}\pi_x w_{xy}(e^{ra_{xy}} - 1) = \sum_{x\neq y}\bar{k} f_1(a_{xy})(e^{ra_{xy}} - 1) = \bar{k}[\alpha_+ \sum^+_{x\neq y} f_{1+}(a_{xy})e^{ra_{xy}} + \alpha_- \sum^-_{x\neq y} f_{1-}(a_{xy})e^{ra_{xy}} - 1]; \tag{A.3}$$

we get $\bar{k}\alpha_+ = \Sigma^+_{x\neq y}\pi_x w_{xy}$, $\bar{k}\alpha_- = \Sigma^-_{x\neq y}\pi_x w_{xy}$, where $\alpha_+ + \alpha_- = 1$, $\alpha_+$ is share of values $a_{xy} > 0$, $\alpha_-$ is share of values $a_{xy} < 0$; $\Sigma^+$ includes quantities $a_{xy} > 0$, $\Sigma^-$ includes quantities $a_{xy} < 0$. Let's move from the sum in Eq. (A.3) to the integrals for positive and negative values $a_{xy}$. For $a_{xy} < 0$,

$\sum^- = \sum^-_{x\neq y}\bar{k} f_{1-}(a_{xy})(e^{ra_{xy}} - 1) \approx \int_{-\infty}^0 \bar{k} f_{1-}(x)(e^{rx} - 1)dx$.

Making the substitution $x \to -x$, we get that:

$\sum^- \approx \int_0^\infty \bar{k} f_{2-}(x)(e^{-rx} - 1)dx = -\bar{k}r/(m_{3-} + r),\quad f_{2-}(x) = m_{3-}e^{-m_{3-}x},\ x > 0,\quad m_{3-} = \bar{k}/\left|\langle a\rangle_{\pi-}\right|,\ \langle a\rangle_{\pi-} = \langle a\rangle_\pi\big|_{a_{xy}<0}$.

We perform the same operation with the term $\sum^+$, obtaining $\sum^+ \approx \bar{k}r/(m_{3+} - r)$. We write the first term in Eq. (A1) as $\dfrac{\lambda_1 r^2}{c_1 - r} = \dfrac{\lambda_1 c_1 r}{c_1 - r} - \lambda_1 r$. Then expression (A.1) is rewritten as:

$$k(r) = -\frac{\alpha_- \bar{k} r}{m_{3-} + r} + \frac{\alpha_+ \bar{k} r}{m_{3+} - r} + \frac{\lambda_1 c_1 r}{c_1 - r} - \lambda_1 r. \tag{A.4}$$

Expression (A.4) was obtained using Eqs. (A.2)-(A.3). Value:

$$m = \partial k(r)/\partial r\big|_{r=0} = \langle a\rangle_\pi = \alpha_+\langle a\rangle_{\pi+} + \alpha_-\langle a\rangle_{\pi-} = \alpha_+\langle a\rangle_{\pi+} - \alpha_-\left|\langle a\rangle_{\pi-}\right|. \tag{A.5}$$

Consider the case $m$<0. Theorem 3.5 (p. 120) work [57] shows that for $m$<0,

$$\rho_{+|s\to 0} \to \rho_{+} = c_1 p_{+} > 0\,, \tag{A.6}$$

where $\rho_+$ is the positive root $r_{s0}$ of the equation $k(r_{s0})$=0, $P\{\xi^+ > x\} = q_+ e^{-x\rho_+}$, $x > 0$, $q_+ = 1 - p_+$, $p_+ = \int_0^\infty P\{-\theta'_c \le \xi(t) < 0\}dt \,/ \int_0^\infty P\{\xi(t) \ge -\theta'_c\}dt$, $\xi^+$ is the maximum of the process $\xi(t)$ (in our case the process Eq. (1)), the parameter $\theta'_c$ is distributed according to: $P\{\theta'_c > t\} = e^{-ct}$, $t \ge 0$.

To model the cumulant (4), as was done using the exponential distribution in expressions (26), we divide the sum in expression (4) into two terms with positive and negative values of the parameter $a_{xy}$:

$$\Sigma = \sum\nolimits_{x\ne y} \pi_x w_{xy} e^{-sa_{xy}} = \sum\nolimits_{x\ne y}^{+} \pi_x w_{xy} e^{-sa_{xy}} + \sum\nolimits_{x\ne y}^{-} \pi_x w_{xy} e^{-sa_{xy}}\,, \tag{A.7}$$

where $\Sigma^+$ includes the quantities $a_{xy} > 0$, and $\Sigma^-$ includes the quantities $a_{xy} < 0$. Having designated:

$$f_+ = \frac{\pi_x w_{xy}}{\alpha_+ \Sigma_{x\ne y}\pi_x w_{xy}}, \quad f_+ \in \Sigma^+, \quad f_- = \frac{\pi_x w_{xy}}{\alpha_- \Sigma_{x\ne y}\pi_x w_{xy}}, \quad f_- \in \Sigma^-, \quad \bar{k} = \Sigma_{x\ne y}\pi_x w_{xy}\,, \tag{A.8}$$

Then Eq. (A.7) is rewritten in the form:

$$\Sigma = \bar{k}[\Sigma^+ f_+ \chi_{a_{xy}} e^{-sa_{xy}} + \Sigma^- f_-(1-\chi_{a_{xy}})e^{-sa_{xy}}], \quad \chi_{a_{xy}} = \begin{cases} 1, & a_{xy} > 0 \\ 0, & a_{xy} < 0 \end{cases}.$$

Thus, for the four-state model considered in works [1], [4]:

$$\bar{a}_+ = \frac{24}{4}0.9 = 5.4, \quad \bar{a}_- = -\frac{43}{4}0.9 = -9.675, \quad \bar{a} = \bar{a}_+ + \bar{a}_- = -\frac{19}{4}0.9 = -4.275, \quad \bar{k} = \frac{67}{4} = 16.75$$

at $\pi_i = 1/4, i = 1,2,3,4, \quad \alpha_+ = 0.36, \quad \alpha_- = 0.64$.

The first term of the cumulant from Eq. (4) is written in the form:

$$\Sigma = \lambda_+(e^{-sa_{xy}} - 1) + \lambda_-(e^{-sa_{xy}} - 1), \quad \lambda_\pm = \bar{k}\alpha_\pm\,, \tag{A.9}$$

where the first term Eq. (A.9) includes the quantities $a_{xy} > 0$, and the second $a_{xy} < 0$. Independence $a_{xy}$ from $\pi_x w_{xy}$. Expression (A.9) is the sum of the cumulants of a complex Poisson distribution. Let's write it down:

$$\theta(t) = \lambda[\Psi(t) - 1], \quad \Psi(t) = Ee^{it\xi} = \int_{-\infty}^{\infty} e^{itx} p(x)dx, \quad it = -s, \quad p_\pm = \delta(x - a_{xy\pm}), \; a_+ > 0, \; a_- < 0. \tag{A.10}$$

In expressions (A.9)-(A.10), the characteristic function $\Psi(t)$ of random variables $\xi_k$ from the sum of random variables $\zeta = \sum_{k=1}^{\nu} \xi_k$, where $\xi_1, \xi_2, \ldots$ is a sequence of independent identically distributed random variables and $\nu$ is a random variable subject to the Poisson distribution with parameter $\lambda$ includes a delta distribution of the form $\delta(x - a_{xy\pm})$. It can be assumed that this distribution can be replaced by another distribution, taking into account possible fluctuations of the random variable $\xi_k$ that are absent in the case of a delta distribution. In our case, the distribution $\delta(x - a_{xy\pm})$ describes the situation when the number of parameters $a_{xy}$ changes with jumps of one.

In such a description, the distributions $f$ from Eq. (A.8) do not appear, but only fluctuations of random variables $\xi_k$ are taken into account. If we return to the distribution of the form Eqs. (A.7)-(A.8) and consider distributions for various types, we obtain a similar description.

With this approach, the probability $p$ remains in the form of a delta-function, but the probability density $f$ is specified from Eq. (A.8). The explicit form of the first term of the cumulant Eq. (4)

was recorded for various distributions: exponential, uniform, $\chi^2$-distribution. The calculation shows that a uniform distribution is closest to the exact behavior for the four-state model.

When $r=x$ the equation for $x$, obtained from Eq. (A.4), is:

$$r_s = x, \quad x^4 + ax^3 + bx^2 + cx + d = 0, \quad a = k_1 + \gamma_1 + m_{3-} - m_{3+}, \quad k_1 = \langle k \rangle / \lambda_1, \quad \gamma_1 = \gamma / \lambda_1, \quad \text{(A.11)}$$

$$b = k_1(\alpha_+ m_{3-} - \alpha_- m_{3+} - c_1) - m_{3+}m_{3-} + \gamma_1(m_{3-} - m_{3+} + c_1), \quad c = k_1 c_1(\alpha_- m_{3+} - \alpha_+ m_{3-}) + 2c_1 m_{3+} + m_{3+}m_{3-},$$

$$d = \gamma_1 m_{3+} m_{3-} c_1 .$$

Solutions of equation (A11) have the form $x_i = y_i - a/3$,

$$y_{1,2} = \sqrt{2s-p}/2 \pm \sqrt{(2s-p)/4 - s - q/2\sqrt{2s-p}}, \quad y_{3,4} = -\sqrt{2s-p}/2 \pm \sqrt{(2s-p)/4 - s + q/2\sqrt{2s-p}}, \quad \text{(A.12)}$$

where for $s$ we obtain a third-order equation

$$a_3 s^3 + b_3 s^2 + c_3 s + d_3 = 0, \quad a_3 = 2, b_3 = -p, c_3 = -2r, d_{3-} = rp - q^2/4$$

with solutions:

$$s = y_{31} + p/6, \quad p_3 = (3a_3c_3 - b^2{}_3)/3a_3{}^2, \quad q_3 = (27a^2{}_3 d_3 - 9a_3b_3c_3 + 2b_3{}^3)/27a_3{}^3,$$

$$y_{31} = \sqrt[3]{-\frac{q_3}{2} + \sqrt{(\frac{q_3}{2})^2 + (\frac{p_3}{3})^3}} + \sqrt[3]{-\frac{q_3}{2} - \sqrt{(\frac{q_3}{2})^2 + (\frac{p_3}{3})^3}}, \quad y_{32,33} = -\frac{y_{31}}{2} \pm \sqrt{(\frac{y_{31}}{2})^2 + (\frac{q_3}{y_{31}})} .$$

The relation (40) includes the derivative $\partial\rho_+ / \partial\gamma$. When $x_1=\rho$,

$$\frac{\partial\rho_+}{\partial\gamma} = \frac{\partial y_1}{\partial\gamma} - \frac{1}{3}\frac{\partial a}{\partial\gamma}, \quad \frac{\partial a}{\partial\gamma} = \frac{1}{\lambda_1}, \quad \frac{\partial c}{\partial\gamma} = 0, \quad \frac{\partial b}{\partial\gamma} = \frac{1}{\lambda_1}(m_{3-} - m_{3+} - c_1), \quad \frac{\partial d}{\partial\gamma} = m_{3-}m_{3+}c_1,$$

$$\frac{\partial y_1}{\partial\gamma} = \frac{2\partial s/\partial\gamma - \partial p/\partial\gamma}{4\sqrt{2s-p}} + \frac{(2\partial s/\partial\gamma - \partial p/\partial\gamma)/4 - \partial s/\partial\gamma - (\partial q/\partial\gamma)/2\sqrt{2s-p} + q(2\partial s/\partial\gamma - \partial p/\partial\gamma)/4(2s-p)^{3/2}}{2\sqrt{(2s-p)/4 - s - q/2\sqrt{2s-p}}},$$

$$\frac{\partial p}{\partial\gamma} = \frac{\partial b}{\partial\gamma} - \frac{3a}{4}\frac{\partial a}{\partial\gamma}, \quad \frac{\partial q}{\partial\gamma} = \frac{3a^2}{8}\frac{\partial a}{\partial\gamma} - \frac{1}{2}(\frac{\partial a}{\partial\gamma}b + a\frac{\partial b}{\partial\gamma}) + \frac{\partial c}{\partial\gamma},$$

$$\frac{\partial r}{\partial\gamma} = -\frac{12a^3}{256}\frac{\partial a}{\partial\gamma} + \frac{1}{16}(\frac{\partial q}{\partial\gamma}2ab + a^2\frac{\partial b}{\partial\gamma}) + \frac{\partial d}{\partial\gamma} - \frac{c}{4}\frac{\partial a}{\partial\gamma},$$

$$\frac{\partial s}{\partial\gamma} = \frac{\partial y_{31}}{\partial\gamma} + \frac{1}{6}\frac{\partial p}{\partial\gamma}, \quad \frac{\partial y_{31}}{\partial\gamma} = \frac{1}{3}[-\frac{q_3}{2} + \sqrt{(\frac{q_3}{2})^2 + (\frac{p_3}{3})^3}]^{-2/3}[-\frac{1}{2}\frac{\partial q_3}{\partial\gamma} + \frac{q_3\partial q_3/\partial\gamma + (p_3/3)^2\partial p_3/\partial\gamma}{2\sqrt{(\frac{q_3}{2})^2 + (\frac{p_3}{3})^3}}],$$

$$p_3 = -\frac{1}{12}(12r + p^2), \quad q_3 = \frac{1}{216}(96rp - 2p^3 - 27q^2),$$

$$\frac{\partial q_3}{\partial\gamma} = \frac{1}{216}[96(\frac{\partial r}{\partial\gamma}p + \frac{\partial p}{\partial\gamma}r) - 54q\frac{\partial q}{\partial\gamma} - p^2\frac{\partial p}{\partial\gamma}], \quad \frac{\partial p_3}{\partial\gamma} = -\frac{1}{12}[12\frac{\partial r}{\partial\gamma} - 2p\frac{\partial p}{\partial\gamma}].$$